\documentclass[11pt, onehalfspacing]{article}

\usepackage{fullpage} 
\usepackage{setspace}	
\usepackage{authblk}
\usepackage[font=small,labelfont=bf]{caption}

\usepackage{amssymb}
\usepackage{amsmath}
\usepackage{graphics, graphicx}
\usepackage{bm}
\usepackage{color,url}
\usepackage{natbib}
\usepackage{algorithm}
\usepackage[noend]{algpseudocode}
\usepackage{float}

\newcommand{\brm}[1]{\bm{\mathrm{#1}}}
\newcommand{\ang}{\AA ngstr\"{o}ms}

\newcommand{\tp}{\intercal}

\newcommand{\cnt}{y_{ij}}
\newcommand{\cntk}{y_{ijk}}
\newcommand{\Cnt}{Y_{ij}}

\newcommand{\tm}{t_j}
\newcommand{\wl}{w_i}
\newcommand{\lam}{\lambda(\tm, \wl)}
\newcommand{\lamnot}{\lambda(\wl)}
\newcommand{\lambnot}{\lambda_b(\wl)}
\newcommand{\boldbeta}{\bm{\beta}}
\newcommand{\boldeta}{\bm{\eta}}
\newcommand{\boldtheta}{\bm{\theta}}
\newcommand{\pplike}{\log{\cal L}_{\mathsf{pp}}}
\newcommand{\ppplike}{\log {\cal L}_{\mathsf{ppp}}}
\newcommand{\fkcom}{FK\,Com}

\newcommand{\ind}{\buildrel{\rm ind}\over\sim}

\newcommand{\ignore}[1]{}

\newcommand{\changes}[1]{#1}

\begin{document}

\title{\bf Detecting Abrupt Changes in the Spectra of High-Energy Astrophysical Sources}

\author[1]{Raymond~K.~W.~Wong\thanks{\noindent Raymond K. W. Wong is an Assistant Professor in the Statistics Department at Iowa State University. Email: {\tt raywong@iastate.edu}}}
\author[2]{Vinay~L.~Kashyap\thanks{\noindent Vinay L. Kashyap is an Astrophysicist at the Harvard-Smithsonian Center for Astrophysics. Email: {\tt vkashyap@cfa.harvard.edu}}}
\author[3]{Thomas~C.~M.~Lee\thanks{\noindent Thomas C. M. Lee is a Professor in the Statistics Department at the University of California, Davis. Email: {\tt tcmlee@ucdavis.edu}}}
\author[4]{David~A.~van~Dyk\thanks{\noindent David A. van Dyk is a Professor in the Statistics Section at Imperial College London. Email: {\tt dvandyk@imperial.ac.uk}}}
\affil[1]{Department of Statistics, Iowa State University}
\affil[2]{Harvard-Smithsonian Center for Astrophysics}
\affil[3]{Department of Statistics, University of California, Davis}
\affil[4]{Statistics Section, Department of Mathematics, Imperial College London}

\date{December 10, 2015}

\maketitle

\begin{abstract}

Variable-intensity astronomical sources are the result of complex and often extreme physical processes. 
Abrupt changes in source intensity are typically accompanied by equally sudden spectral shifts, i.e., sudden changes in the wavelength distribution of the emission.  This article develops a method for modeling photon counts collected \changes{from} observation of such sources. We embed change points into a marked Poisson process, where photon wavelengths are regarded as marks and both the Poisson intensity parameter and the distribution of the marks are allowed to change.  \changes{To the best of our knowledge} this is the first effort to embed change points into a {\it marked} Poisson process. Between the change points, the spectrum is modeled non-parametrically using a mixture of a smooth radial basis expansion and a number of local deviations from the smooth term representing spectral emission lines. Because the model is over parameterized we employ an $\ell_1$ penalty.  The tuning parameter in the penalty and the number of change points are determined via the minimum description length principle. Our method is validated via a series of simulation studies and its practical utility is illustrated in the analysis of  the ultra-fast rotating yellow giant star known as \fkcom. 

\end{abstract}

\section{Introduction}
\label{sec:intro}
Astronomical sources that exhibit temporal variability or periodicity are among the most scientifically rich in the observed Universe.  Variation in source intensity can result from rotations, eclipses, magnetic activity cycles, or the turbulent flows of matter into deep gravitational wells.  Massive stellar explosions known as {\it super novae}, for example,  appear as abrupt peaks in the electromagnetic radiation emitted from a source. They sometime produce a narrow beam of intense high-energy radiation that appears as $\gamma$-ray bursts from Earth---although there are other sources of $\gamma$-ray bursts. These bursts originate in distant galaxies, can last from milliseconds to minutes, and are among the most energetic events that occur in the Universe. There are many less dramatic objects that produce temporal variability; they include signals with repeated peaks and troughs that may or may not follow a predictable periodic pattern. A {\it variable star}, for example, exhibits fluctuating emission that can be due to a natural expansion and contraction in its radius as it evolves (a so-called {\it pulsating variable star}) or to eruptions in its coronae such as flares or mass ejections. In extreme cases, superflares can erupt producing millions of times more energy than a typical solar flare; if such a flare occured on the Sun it could destroy the Earth's ozone layer and cause mass planetary extinction.  Although the mechanism is not well understood, super flares are most likely to occur on fast rotating stars \citep{mcke:12}. 
  
In this article we develop statistical methods that enable us to identify changes in observed emission from astronomical sources, such as those associated with massive energetic events in the coronae of variable stars. Dramatic shifts in intensity of these sources are typically accompanied by changes in the distribution of the wavelength of the photons emitted, known as the {\it spectrum} of the source. Thus, our methods aim to take advantage of changes both in the intensity of the emission and in its wavelength distribution.  We illustrate our methods by applying them to high-energy observations of the flaring star known as  \fkcom. 
 
 \paragraph{\it Data collection in high-energy astrophysics} 
High-energy astrophysics is the study of electromagnetic radiation in the ultraviolet, X-ray, and $\gamma$-ray bands. The production of such high-energy photons requires temperatures of millions of degrees and signals the release of deep wells of stored energy such as those in very strong magnetic fields, extreme gravity, explosive nuclear forces, and shock waves in hot plasmas. High-energy detectors have much in common in terms of their statistical properties (and much that distinguishes them from detectors in other energy regimes). In this article we focus on statistical methods that are appropriate for high-energy astrophysics. 

Astronomical data from high-energy observatories are usually obtained as a list of photons;  the list records the two-dimensional direction from whence each photon arrived, the time at which it was recorded, and its wavelength (and hence its energy).  \changes{Owing to the intrinsic resolutions of high-energy photon telescope/detector systems, each of these four attributes is inherently discrete and the data can be compiled into a four-way table of photon counts.}  Although each attribute is subject to uncertainty (also inherent in the telescope/detector systems, e.g., the true direction of arrival is distorted by the point spread function), the quality of the resulting datasets is unprecedented in the history of astronomical observations.  Nonetheless the remarkable four-way tables of photon attributes are largely an untapped resource.  This is mostly due to a lack of principled statistical methods that can be used to detect and characterize astronomical sources in high-dimensional spaces.  This is especially true of high-spectral-resolution grating data (e.g., data obtained using the HETGS+ACIS-S configuration of the {\it Chandra X-ray Observatory}), where a self-selected sample of {\sl interesting} objects are observed for long durations (often for $>100$~ksec) with Doppler resolutions of ${\sim}100$~km~s$^{-1}$, and with an intrinsic temporal resolution that varies from one millisecond to three seconds.
Because the locations of the objects we study in this article do not change,  we ignore directional information and focus on the two-way table of photon counts indexed by time and wavelength. The one-way marginal table of wavelengths is called the (observed) {\it spectrum} and that of the times is the (observed) {\it light curve} of the source. 

\paragraph{\it Change points in Poisson processes}
From a statistical point of view, sudden changes in an underlying physical process are modeled via change points. In a parametric model, for example, the value of the parameter is allowed to change at one or more time points during the observation. The periods of time where the parameter is fixed are called the {\it regimes} of the process. We might consider testing for a single change point, fitting the model with a given number of change points, or estimating the number of change points.  Given that the data in high-energy astrophysics are photon counts, we focus on change points in non-homogeneous Poisson processes. Because the wavelength (and direction) \changes{of} each photon are recorded, we can either model them using a marked Poisson process or count the events in each of a number of wavelength bins and model them as a multivariate Poisson process. Typical detectors have discrete temporal and spectral resolution so it is natural to use photon counts in time cross wavelength bins, i.e., to use discrete-time processes.
\changes{The methods that we propose are designed specifically for two-way tables of Poisson counts of this sort.}

The complexity of any statistical model naturally increases with the number of change points. Thus to avoid overfitting, methods must penalize complexity. Bayesian prior distributions and marginalization techniques are natural avenues for encouraging parsimony, for example in that they inherently embody Occam's Razor in model selection  \citep[e.g.,][chapter 28]{mack:03}. It is not surprising then that there is an extensive literature on Bayesian methods for change point models. \citet{raft:akma:86} is an early contribution for count data generated according to a homogeneous Poisson process until some unknown time when \changes{the} intensity of the process changes.  They propose computing the joint posterior distribution of the two Poisson rates and the single change point along with a Bayes Factor to test for existence of the change point; see \citet{carl:etal:92}  and \citet{more:etal:05} for derivation of hierarchal and intrinsic prior distributions, respectively, in the single change point Poisson setting.  \citet{gree:95} generalizes this approach by using a continuous time model, allowing for multiple regimes, putting a prior distribution on the number of regimes, and fitting this number.  \citet{lai:xing:11}, on the other hand, use a discrete time model with a Bernoulli process governing the change in regimes. This approach follows \citet{chib:98} who embedded a latent discrete-time discrete-state Markov process for the regime index into a multi-level Bayesian model. This formulation has been applied for inhomogeneous Poisson processes, for example, to model change points in the parameters of a log-linear model \citep{park:10}, see \citet{park:etal:12} for a related approach.  

Perhaps the most popular change-point method for astronomical count data is known as Bayesian Blocks \citep{scar:98}. It starts by splitting the time interval into two, assuming a homogeneous Poisson process on each, and computing the posterior odds of this model relative to a homogeneous process on the entire interval.  The change point is chosen by maximizing the posterior odds and is accepted if the odds favor this model. This process continues recursively on each subinterval until no more change points are added. \citet{scar:etal:13} improved the method by achieving global optimization over the change points while adding the capacity to handle gaps in the data,  variable exposure, piecewise linear and piecewise exponential Poisson intensities, multivariate data, the analysis of variance,  data on the circle, etc. Bayesian Blocks does not, however, address changes in the spectrum at the change points. 

Non-Bayesian methods for change-point detection have also been proposed for non-homogeneous Poisson processes.  \citet{Akman-Raftery86} and \citet{Worsley86} are early contributions.  Both papers developed methods for testing for the presence of a change-point in a Poisson process.  These methods follow classical frequentist arguments and also provide interval estimates for any detected change-points.  \citet{Loader92} considered using log-linear models for non-homogeneous Poisson processes.  In particular likelihood ratio tests are derived to choose a ``best'' model amongst different change-point and log-linear models for the observed data.  In \citet{Mei-et-al11} methods are proposed for detecting changes in Poisson rate after the effects of population size are taken into account.  These methods are based on a generalized likelihood ratio, weighted likelihood ratio, and adaptive thresholding.  More recently, \citet{shen:zhan:12} applied the modified BIC (mBIC) of \citet{Zhang-Siegmund07} to detect change-points in non-homogeneous Poisson processes.  
\changes{
The mBIC can be viewed as a penalized likelihood model selection criterion that is tailored to estimating the number and locations of the change points.
}

\paragraph{\it Change points in marked Poisson processes} 
Despite the extensive literature on change points in Poisson processes,
there is little on change points in {\it marked} Poisson processes.
\changes{Methods  have been developed to test for non-homogeneity in marked Poisson processes, for example via a scan statistic that is computed in a sliding time interval \citep{chan:zhan:07}.
}
 In high-energy astrophysics a common strategy is to compute a simple summary statistic describing the wavelength distribution in each time bin and to visually inspect a plot of the statistic as a function of time. A typical choice of statistics is the so-called {\it hardness ratio} which is the ratio of the photon count observed above a given wavelength threshold to that below the threshold; see \citet{park:etal:06} for a small-sample statistical treatment of hardness ratios. Plotting hardness ratios against time is an informal exploratory technique. The primary goal of this article is to provide a change point method for marked Poisson processes that is inspired by and tailored to a specific problem in high-energy astrophysics.

\paragraph{\it \changes{Methods for detecting change points}}
\changes{
There is a wide range of  statistical approaches for detecting statistical change points.  For the classical simple case of piecewise constant signals with additive noise, for example, \citet{Yao88} studied the use of the Schwartz criterion and showed it to be consistent.  This criterion can be interpreted as the classical Bayesian information criterion (BIC) if the location of a change point is treated as a free parameter (or as a dimension of the model). In this case, however, the likelihood functions violate the regularity conditions necessary for the classical BIC, so both criteria can likely be improved. To circumvent this issue, \citet{Zhang-Siegmund07} provide an ingenious method to approximate the Bayes factor and developed the above-mentioned mBIC for change point detection which has been shown to be consistent in various settings.  More recently, \citet{Ninomiya14} derived a specialized Akaike information criterion (AIC) for change point estimation.  Results from \citet[][Section~3.3]{Aue-Lee11}, however, can be used to show that this AIC method is not consistent.

In this article we focus on the minimum description length (MDL) principle \citep{Rissanen89,Rissanen07}, which can also be viewed as a penalized likelihood approach to model selection. A brief introduction of MDL is given in Section \ref{sec:firstMDL}. One of its earliest applications to change point estimation is the image segmentation work of \citet{Leclerc89}, where the corresponding MDL criterion can be easily reduced to the classical one-dimensional change point problem setting; see also \citet{Lee98:segcor} for an extension to correlated noise.  Other successful MDL applications to change point problems include discontinuity detection in nonparametric regression \citep{Lee02}, structural break detection in nonstationary time series models \citep{Davis-Yau13,Yau-et-al15}, and change point detection in quantile modeling \citep{Aue-et-al14b}.  Under mild regularity assumptions, these MDL solutions can be shown to be consistent.

For many change-point problems simpler than the one considered in this paper, the major penalty terms of mBIC and MDL are remarkably similar; e.g., compare Equation~(8) of \citet{Zhang-Siegmund07} and Equation~(6.2) of \citet{Lee97c}.  This provides reassuring evidence that both mBIC and MDL are reliable methods for change point detection.}

\paragraph{\it \changes{Strategy and outline}} We pursue a non-parametric penalized-likelihood strategy, using MDL to optimize the tuning parameters of the penalty function and to fit the number of change points.  Between the change points, we assume a homogeneous Poisson process where the spectrum includes a broad smooth term that is modeled using a radial basis expansion.  In particular the bases we use are cubic polynomials, which provide a good balance between model flexibility and ease of implementation.  We also include a number of single-bin local deviations from this extended smooth model. Physically, these deviations correspond to spectral emission lines that are the result of the production of photons of a particular wavelength as electrons fall to the lower energy shells of a specific ion. Both the radical basis expansion and the local deviations are over parameterized in that we expect most of their coefficients and intensities (respectively) to be zero. Thus, we employ an $\ell_1$ penalty on both.
An R package ``Automark" implementing our proposed methodology
is available to interested readers at
{\tt https://github.com/astrostat/Automark}.

Some simple spectral models such as power laws and exponential absorption can be formulated as log-linear models \citep{vand:etal:01} and thus in principle could be embedded into a temporal model with change points for their parameters, similar to the proposal of \citet{park:10}.  We take a different approach by using flexible non-parametric spectral models. This allows us to account for both blurring of the recorded energies and background contamination of the photon counts, and to apply a single model irrespective of the particular shape of the source spectra. (See \citet{vand:kang:04} for a review of the source and data-distortion models used in high-energy spectral analysis.)  This strategy is contrary to the typical approach in high-energy spectral analysis where physics-based parametric models are 
preferred because their parameters have direct scientific meaning. Our primary goal here, however, is to identify the change points. The spectra corresponding to each regime within the process can be fit using physics-based models in a secondary analysis.

This article is organized into seven sections. Section~\ref{sec:back} describes the detectors used for data collection, the relevant astrophysical models, and basic notation for our statistical approach. We formalize our model  for the homogeneous Poisson processes within each regime of the overall process in Section~\ref{sec:hom} and describe how we embed this into the change-point model in Section~\ref{sec:break}. Numerical results including a set of simulation studies and an application to the yellow giant \fkcom\ appear in Sections~\ref{sec:sim} and \ref{sec:real}, respectively. Discussion appears in Section~\ref{sec:conc} and details of the derivation of our MDL criterion in an Appendix. 

\section{Astrophysical Background and Statistical Notation}
\label{sec:back}

Suppose that we observe photon counts in an $N\times J$ rectangular array of equally-sized wavelength-cross-time bins, where $\cnt$ is the photon count in the bin with wavelength range $[\wl-\delta_w/2, \wl+ \delta_w/2)$ and
temporal range $[\tm-\delta_t/2, \tm+\delta_t/2)$, for $i=1,\ldots, N$ and $j=1,\ldots, J$. \changes{Although this array may be compiled using the natural resolution of the detector, with high-resolution data it may be computationally advantageous to combine the natural bins in order to obtain a lower resolution array of photon counts. This strategy is illustrated and discussed in Section~\ref{sec:real}.}
In any case, the expected count in each bin is ideally proportional to the brightness of the astronomical source integrated over its time and wavelength ranges. Because of detector effects, however, a photon with wavelength $w$ has a distribution of possible recorded energies, $w'$. This distribution is called the {\it detector response function} and denoted $R(w,w')$. We apply our methods to data collected using a  {\it grating} that, like a prism,  spatially separates photons as a function of their wavelength. With grating data, wavelength is measured very accurately and we can ignore measurement uncertainty and treat $R(w,w')$ as a Dirac delta function. Another detector effect arises because the sensitivity of the detector varies with photon wavelength, an effect that is quantified via the so-called {\it effective area} of the detector and denoted by $A(w)$. The effective area quantifies characteristics of the telescope mirrors that do not relate to the nature of the source and is typically treated as known. (See, however, \citet{lee:etal:11} and \citet{xu:etal:14} for development of Bayesian statistical methods that account for uncertainty in $A(w)$.)

Using a Poisson model for the photon counts and letting $\lam$ be the expected count per unit time and per unit wavelength averaged over the bin centered at $(\tm,\wl)$, 

\begin{equation}
\cnt \ind \mathrm{Poisson}\{ \delta_t \cdot  \delta_w  \cdot \lam \cdot A(\wl)\}.
\label{eq:pois-mdl}
\end{equation}
In practice, we may combine data from $K$ detectors, each with its own effective area. Letting  $A_1(w), \ldots, A_K(w)$ denote the $K$ effective areas, the total counts across the $K$ detectors can be modeled
\[
\Cnt = \sum_{k=1}^K \cntk,
\]
where
\[
\cntk \ind \mathrm{Poisson}\{ \delta_t \cdot \delta_w \cdot \lam \cdot A_k(\wl)\}
\]
are the counts \changes{recorded with} detector $k$.  Because the $\cntk$ are independent  across $k$, we have
\[
\Cnt \sim \mathrm{Poisson}\left\{  \delta_t \cdot \delta_w \cdot \lam \sum_{k=1}^K A_k(\wl)\right\}.
\]

The goal of this article is to infer the properties of $\lam$ using the observed photon counts, $\{\Cnt\}$.  We are particularly interested in detecting statistically significant changes in $\{\Cnt\}$ over time.  In other words, we would like to determine if there are any change point, $\pi$, such that $\lam\big|_{\{\tm\leq \pi\}}$ \changes{differs} from $\lam\big|_{\{\tm>\pi\}}$.  If there are, we aim to estimate the number of change points and their values.  

For ease of presentation, we develop our estimation procedure in two stages.  First  we develop a flexible model for a time-homogeneous source spectrum. That is we assume that there are no change points and that $\lam$ is constant in $\tm$.
In the second stage we  introduce change points and allow $\lam$ to change over time.

\section{A Model for the Homogeneous Poisson Processes}
\label{sec:hom}

\subsection{A nonparametric spectral model}
Because there are no change points in our time-homogeneous Poisson model, we 
drop $\tm$ from the argument of $\lam$ and write $\lam=\lamnot$
in this section. Also, we denote the expectation of $Y_{ij}$ by $\mu(w_i)$.
That is, $\mu(w_i) = s(w_i) \lambda(w_i)$ with
$s(w_i) = \delta_t\cdot\delta_w\sum^K_{k=1}A_k(w_i)$ being a completely specified function.
We represent the expected counts due to the underlying physical process by $\lambda$ and  expected detector counts by $\mu$; $\mu$ is adjusted for varying bin sizes and effective areas.  

The parameter of scientific interest is $\lamnot$. For the astronomical sources \changes{that} we study $\lamnot$ is mostly smooth with a few emission lines overlaid and we adopt the model
\begin{equation}
\lamnot = f(\wl)+\sum^M_{m=1}\alpha_m g_m(\wl,\tau_m), \label{fun:lambdaw}
\end{equation}
where $f(w)$ is a nonnegative smooth function, $M$ is the number of emission lines, $\alpha_m > 0$ is the magnitude of emission line $m$, and $g_m(w,\tau_m)$ models the shape of the emission line centered at $\tau_m$.
Common choices for $g_m(w,\tau_m)$ include Gaussian density functions with small
variances, Lorentzian density functions, and delta functions 
\citep[e.g.,][and the references therein]{Dyk-Connors-Esch06}.
Here we assume that each of the emission lines locates completely within a single bin, although our framework can be easily modified for cases when some emission lines span multiple bins.
For astrophysical reasons, we also assume that each $\alpha_m >0$, that is each emission line constitutes a positive deviation from $f(w)$, although in principle  the $\alpha_m$ may also be negative, so long as $\lamnot>0$.

If an individual spectral line is located in a single bin, the line profile, $g_m(w, \tau_m)$, is an indicator function for the interval of width $\delta_w$ centered at $\tau_m\in\{w_1,\dots,w_N\}$, that is, an indicator function for one of the wavelength bins. The natural resolution of the bin counts does not allow modeling of  $g_m(w, \tau_m)$ at any greater level of detail than this.
Under this model for the emission lines,  $\lamnot$ in (\ref{fun:lambdaw}) becomes
\[
\lambda(\wl) =
 f(\wl) + \sum^M_{m=1}\alpha_m I\left(\tau_m-\frac{\delta_w}{2} \le \wl < \tau_m+\frac{\delta_w}{2}\right),
\]
where
 $I(\mathcal{A})$ denotes the indicator function for $\mathcal{A}$ and  $\tau_m\in\{w_1,\dots,w_N\}$.
Because of the nonnegativity constraint on $\lambda(\wl)$, we introduce a log-link function,
\begin{equation}
  \log\lambda(\wl) =  \log f(\wl) + \sum^M_{m=1}\alpha_m' I\left(\tau_m-\frac{\delta_w}{2} \le \wl < \tau_m+\frac{\delta_w}{2}\right) 
 \label{eq:log-link} 
\end{equation}
 where $\alpha_m'$ is the transformed  $\alpha_m$ that ensures the equality in (\ref{eq:log-link}) holds.

To provide flexible modeling, we use a radial basis expansion \changes{\citep[see, e.g.,][]{Buhmann03, Ruppert-Wand-Carroll03}} to nonparametrically model $\log f(w)$, the smooth component of $\log\lambda(w)$.  Specifically, we use the $P$ basis functions, $(\xi_1(w), \ldots, \xi_P(w))$, which are polynomials of power 3 with \changes{equally spaced} knots, $\kappa_1, \ldots, \kappa_{P-4}$.\footnote{\changes{The gap between the first knot the left endpoint of the spectral range and the gap between the last knot and the right endpoint of the range are both slightly larger than the gaps between consecutive knots.}
}
\changes{Specifically, the basis functions are}
\begin{align*}
&\xi_1(w) = 1,\quad \xi_2(w) = w, \quad \xi_3(w)=w^2,\quad \xi_4(w)=w^3,\quad \mbox{and}\\
&\xi_{p+4}(w) = |w -\kappa_p|^3, \quad\quad p=1,\dots,P-4.
\end{align*}
\changes{
Substituting these into (\ref{eq:log-link}) yields}
\begin{align}
\log\lamnot  & = 
  \sum_{p=1}^{P}\beta_p \xi_p(\wl) + \sum^M_{m=1}\alpha_m' I\left(\tau_m-\frac{\delta_w}{2} \le \wl < \tau_m+\frac{\delta_w}{2}\right),
\label{eqn:gensemi}
\end{align}
where
$\beta_p$ is the coefficient of basis function $p$. 

Model~(\ref{eqn:gensemi}) is an example of a generalized semi-parametric model. Under this formulation, estimating $\lamnot$ is equivalent to estimating $P$, $(\beta_1, \ldots, \beta_P)$, $M$, and $(\alpha_1', \tau_1, \ldots, \alpha_M', \tau_M)$.

\subsection{A lasso model for the emission lines}
A difficulty in fitting model~(\ref{eqn:gensemi}) is the estimation of $(\tau_1, \ldots, \tau_M)$, the locations of the emission lines.  With $M$  unknown, there is a total of $2^N$ possibilities for $(\tau_1,\ldots,  \tau_M)$.  However, as emission lines are relatively rare (i.e., sparse), we can take advantage of the celebrated $\ell_1$ penalty techniques \citep[e.g.,][]{Tibshirani96} to provide a fast searching algorithm for $(\tau_1, \ldots, \tau_M)$.

We begin by rewriting~(\ref{eqn:gensemi}) as
\begin{equation}
  \log\lamnot 
  = \sum^P_{p=1}\beta_p \xi_p(\wl) + \sum^N_{k=1} \eta_k I_{k}(\wl)  
  = \sum^P_{p=1}\beta_p \xi_p(\wl) +\eta_i,
\label{eqn:pmodel}
\end{equation}
with $I_{k}(w) = I(w_k-\delta_w/2 \le w< w_k+\delta_w/2)$.  In this formulation $\eta_i=0$ implies there is no emission line in the wavelength bin centered at $w_i$.  There is a one-to-one correspondence between models~(\ref{eqn:gensemi}) and~(\ref{eqn:pmodel}) via 
\begin{align*}
M = \sum^N_{i=1} I(\eta_i\neq 0) \quad \mbox{and} \quad
\Big\{(\tau_m,\alpha_m')\Big\}^M_{m=1} = \Big\{(w_i, \eta_i): \eta_i\neq0\Big\}.
\end{align*}
A major advantage of re-expressing model~(\ref{eqn:gensemi}) as (\ref{eqn:pmodel}) is that, many of the $\boldeta=(\eta_1,\ldots, \eta_N)^\tp$ are zero in~(\ref{eqn:pmodel}). Therefore by imposing an $\ell_1$ penalty, we can achieve fast simultaneous selection and estimation of the nonzero elements among $\boldeta$.

We employ a common strategy for the smooth component, $\log f(\wl)=\sum_{p=1}^P \beta_p \xi_p(\wl)$, in~(\ref{eqn:pmodel}).  Namely, we specify a large value of $P$ {\it a priori} and estimate $\boldbeta=(\beta_1, \ldots, \beta_P)^\tp$ under a penalty term that aims to avoid over-fitting.  In this way, the coefficients of the the redundant functions among $\xi_1(w), \ldots, \xi_P(w)$ are shrunk toward zero.  This enables us to assume that $P$ is known (and large).

Penalized maximum likelihood can be used to simultaneously estimate all of the unknown parameters in~(\ref{eqn:pmodel}), including those describing the smooth component and those describing the emission lines.

For a Poisson random variable with expectation $a>0$, we write the corresponding
probability mass function as 
\begin{equation}
q(x;a)=-a + x\log a -\log(x!)
\label{eq:pois-loglike}
\end{equation} 
for any
nonnegative integer $x$.
The log-likelihood of a bin count $Y_{ij}$ is 
$$
l_{\mathrm{one}}(Y_{ij}; \bm{\theta}) = q\{Y_{ij}; \mu(w_i)\} = q\{Y_{ij}; s(w_i) \lambda(w_i)\},
$$ 
where $\lamnot$ is parameterized in terms of $\boldtheta=(\boldbeta,\boldeta)$ as in (\ref{eqn:pmodel}).
(This likelihood will be used within one homogenous time segement in Section~\ref{sec:break}; hence the subscript ``one".)
Adding a penalty, we define the estimate of $\boldtheta$ as the maximizer of 
\begin{equation}
\sum^N_{i=1}\sum^J_{j=1}l_{\mathrm{one}}(Y_{ij}; \bm{\theta}) - \gamma \{\rho\|\brm{D}\bm{\beta}\|_1 +
(1-\rho) \|\bm{\eta}\|_1\}, \label{eqn:penlike}
\end{equation}
where $\brm{D}=\mathrm{diag}\{(0,0,0,0,1,\dots,1)\}$ is a diagonal matrix, and $\gamma>0$ and $0\le\rho\le1$ are tuning parameters that determine the penalty on $\bm{\theta}$. {The first four diagonal elements of $\brm D$ are set to zero because there is no penalty for the inclusion of $(\beta_1, \ldots, \beta_4)$ in the model.}   If $\gamma$ and $\rho$ are specified, maximization of~(\ref{eqn:penlike}) is equivalent to lasso fitting of a generalized linear model. There are fast algorithms for this optimization problem \citep[e.g., coordinate descent,][]{Friedman-Hastie-Tibshirani10} and software is widely available; e.g., the $R$ package {\ttfamily glmnet}.

\subsection{Tuning parameter selection using MDL}
\label{sec:firstMDL}
The success of~(\ref{eqn:penlike}) depends heavily on our ability to choose good values of $\gamma$ and $\rho$, which determine the number of emission lines and the number of basis functions used in the smooth component of  the fitted model. To select $\gamma$ and $\rho$, we use the minimum description length (MDL) principle \citep{Rissanen89,Rissanen07}.  In short, MDL defines the best model as the one that achieves the highest compression rate of the data, $\mathcal{D}$.  In other words, the best model allows us to store  $\mathcal{D}$  with the shortest codelength.  A good statistical model and a good compression method share a common feature: they should be able to capture regularities and patterns hidden in the data.  Therefore it is reasonable to expect that a model chosen by MDL  to be a good statistical model; 
\changes{some successful examples for change point problems were provided in Section~\ref{sec:intro}, while for other applications can be found in the later chapters of \citet{grunwald2005advances}.  There are various versions of MDL.}
We  use the so-called two-part form of MDL to derive a model selection criterion for choosing $\gamma$ and $\rho$.  

Let $\mathsf{CL}(\mathcal{Z})$ be the codelength of $\mathcal{Z}$: one can treat $\mathsf{CL}(\mathcal{Z})$ as the amount of memory needed to store $\mathcal{Z}$.
Similarly, let $\mathsf{CL}(\mathcal{Z}|\mathcal{X})$ be the codelength of $\mathcal{Z}$
  conditional on $\mathcal{X}$, { that is, the codelength of $\mathcal{Z}$ with knowledge of $\mathcal{X}$.}
Generally speaking, the two-part form of MDL decomposes the codelength of the data $\mathcal{D}$ into two parts
\begin{equation}
  \mathsf{CL}(\mathcal{D}) = \mathsf{CL}(\widehat{\mathcal{M}}) +
  \mathsf{CL}(\mathcal{D}|\widehat{\mathcal{M}}), \label{eqn:cl}
\end{equation}
where the best fitting model $\widehat{\mathcal{M}}$ is defined to be  the minimizer of $\mathsf{CL}(\mathcal{D})$. 
In Appendix~\ref{app:mdl1} it is shown that when fitting model~(\ref{eqn:penlike}), \changes{for large $N$} the two-part MDL (\ref{eqn:cl})
\changes{is asymptotically equal to}
\begin{equation}
  \mathsf{mdl}_{\mathrm{null}}(\gamma, \rho) =
  -\sum^N_{i=1}\sum^J_{j=1}l_{\mathrm{one}}(Y_{ij};\hat{\bm{\theta}}) +
  \frac{1}{2}\|\hat{\bm{\theta}}\|_0\log (NJ) + 
\log {N \choose \|\hat{\bm{\eta}}\|_0}, \label{eqn:mdl_hom}
\end{equation}
where $\hat{\bm{\theta}}=(\hat{\bm{\beta}}, \hat{\bm{\eta}})$ is the maximizer of~(\ref{eqn:penlike}) given $\gamma$ and $\rho$, and the $\ell_0$ norm, $\|\bm{z}\|_0$, denotes the number of nonzero elements in the vector $\bm{z}$.  \changes{We choose the values of $\gamma$ and $\rho$ that jointly maximize (\ref{eqn:mdl_hom}) by evaluating $\mathsf{mdl}_{\mathrm{null}}(\gamma, \rho)$ on a fine grid, using {\tt glmnet} to compute $\hat{\bm{\theta}}(\gamma,\rho)$ on a grid of values of $\gamma$ for each value of $\rho$.}
In many MDL applications, a term similar to the last one in~(\ref{eqn:mdl_hom}) is  negligible and hence omitted.  Here, however, the number of unknown parameters ($N+P$) is not ignorable when compared to the number of observations and hence the last term  in~(\ref{eqn:mdl_hom}) is required.

In summary, for the time-homogeneous model with no change points, the parameter estimate $\hat{\bm{\theta}}=(\hat{\bm{\beta}}, \hat{\bm{\eta}})$ is obtained by maximizing~(\ref{eqn:penlike}) with $\gamma$ and $\rho$ chosen as the joint minimizer of~(\ref{eqn:mdl_hom}).  Finally, the estimate for $\lambda(\wl)$ is calculated as $\hat{\lambda}(\wl)=\exp\{\sum^P_{p=1}\hat{\beta}_p\xi_p(\wl)+\hat{\eta}_i\}$.

\section{Modeling with Change Points}
\label{sec:break}

\subsection{Adding change points to the spectral model}
\label{sec:addchange}
In this section, we allow the spectrum, $\lam$,  to change over time. Hence we reinstate the notation that emphasized the dependence on $\tm$. We model $\lam$ as piecewise constant as a function of $\tm$.  This involves partitioning the entire time interval $[0,T)$ into several subintervals and independently modeling $\lam$ as constant in $\tm$ in each subinterval using a model of the form given in~(\ref{eqn:pmodel}). 
Again let $\Cnt$ be the photon count in the bin centered at $(\wl, \tm)$ and summed over the $K$ detectors. We can formalize the change-point model for $\lam$ as
\begin{equation}
\lam = \sum^B_{b=1} I_b(\tm) \lambnot,
\label{eqn:piece}
\end{equation}
where 
$B\ge 1$ is the number of time (sub)intervals, 
$I_b(\tm)=I( \pi_{b-1} \le \tm < \pi_{b} )$ identifies the time bins in interval $b$, 
$t_1-\delta_t/2=\pi_0 <\dots<\pi_B=t_J + \delta_t/2$ are the change points, 
and $\lambnot$ is the spectrum in time interval $b$. {For ease of discussion, we refer to $\pi_1,\dots, \pi_{B-1}$ as the {\it change points}, excluding the endpoints $\pi_0$ and $\pi_B$\changes{.}  The within time-interval spectra are modeled as in~(\ref{eqn:pmodel}), i.e.,
\[
  \log\lambnot  = \sum^P_{p=1}\beta_{bp} \xi_p(\wl) +\eta_{bi},
\]
where $\boldbeta_b=(\beta_{b1}, \ldots, \beta_{bP})^\tp$ and $\boldeta_b=(\eta_{b1}, \ldots, \eta_{bN})^\tp$ are the  parameters for the time-interval-specific spectra.

There is no gain in allowing the change points, $(\pi_1,\ldots, \pi_{B-1})$, to take values other than
bin breaks if there is no prior knowledge about the change points. 
Further, to maintain sufficient data within each time interval for acceptable estimation, $|\pi_i-\pi_j|$ cannot be too small for any $0\le i,j\le B$.  We require that $|\pi_i-\pi_j|\ge 5\delta_t$ in our numerical illustrations. This ensures that each time interval is at least five bins wide. 

\subsection{Model selection using MDL}
\label{sec:finalMDL}

In order to fit model~(\ref{eqn:piece}), we must estimate the number of time intervals, $B$, the change points $\bm{\pi}=(\pi_1,\dots,\pi_{B-1})^\tp$, and the parameters \changes{for each interval,} $\bm{\Theta}=\{\bm{\theta}_1,\dots,\bm{\theta}_B\}$, where $\boldtheta_b = (\boldbeta_b, \boldeta_b)$.  Once $B$ and $\bm{\pi}$ are specified, however, each $\lambnot$ along with its tuning parameters, $\gamma=\gamma_b$ and $\rho=\rho_b$, can be estimated individually using the method of Section~\ref{sec:hom}.  Thus, we aim to first estimate $B$, $\bm{\pi}$ and then compute $\hat\gamma_b(B, \bm{\pi})$, $\hat\rho_b(B, \bm{\pi})$, and $\hat\boldtheta_b(\hat\gamma_b, \hat\rho_b)$ as in Section~\ref{sec:hom} for $b=1,\ldots, B$.  
This  notation signifies the dependence of 
$\hat\gamma_b$ and $\hat\rho_b$ on $(B, \bm{\pi})$, and the dependence of $\hat\boldtheta_b$ on $(\hat\gamma_b, \hat\rho_b)$.
Similarly,
we also write ${\bm{\Theta}}$ and $\lambda(t_j,w_i)$ as
${\bm{\Theta}}(\bm{\gamma}, \bm{\rho})$ and
$\lambda\left\{t_j,w_i; B, \bm{\pi}, {\bm{\Theta}}(\bm{\gamma}, \bm{\rho})\right\}$,
respectively.

The profile penalized (pp) loglikelihood function for $B$, $\bm{\pi}$  under model~(\ref{eqn:piece}) depends on the tuning parameters $\bm{\gamma}=(\gamma_1,\dots,\gamma_B)$ and $\bm{\rho}=(\rho_1,\dots, \rho_B)$ and can be written
\begin{equation}
\pplike(B, \bm{\pi}; \bm{\gamma}, \bm{\rho}) =
\sum^N_{i=1}\sum^J_{j=1} q\left(Y_{ij}; s(w_i)\lambda\left\{t_j,w_i; B, \bm{\pi}, \widehat{\bm{\Theta}}(\bm{\gamma}, \bm{\rho})\right\}\right),
\label{eq:pplike}
\end{equation}
\changes{with $q$ the Poisson loglikelihood given in (\ref{eq:pois-loglike}).}
Because $\bm{\gamma}$ and $\bm{\rho}$ can be computed as functions of $B$ and $\bm\pi$, we can substitute 
$\hat{\bm\gamma}(B,\bm\pi)$ and $\hat{\bm\rho}(B,\bm\pi)$ into
$\pplike(B, \bm{\pi}; \bm{\gamma}, \bm{\rho})$.  This is akin to
profiling over $\bm{\gamma}$ and $\bm{\rho}$, except that
$\hat{\bm\gamma}(B,\bm\pi)$ and $\hat{\bm\rho}(B,\bm\pi)$ are computed
according to the MDL criterion rather than by maximizing $\pplike(B,
\bm{\pi}; \bm{\gamma}, \bm{\rho})$. The result is the pseudo profile penalized
(ppp) loglikelihood given by
\begin{align}
  \ppplike(B, \bm{\pi}) &= \pplike(B, \bm{\pi}; \hat{\bm\gamma}(B,\bm\pi), \hat{\bm\rho}(B,\bm\pi)) 
  \nonumber \\
  &= \sum^N_{i=1}\sum^J_{j=1} \sum^B_{b=1} I_b(t_j) l_{\mathrm{one}}\left\{ Y_{ij};\hat\boldtheta_b(B,\bm\pi)  \right\},
\label{eq:llike}
\end{align}
where $\hat\boldtheta_b(B,\bm\pi) = \hat\boldtheta_b\big(\hat{\bm\gamma}(B,\bm\pi), \hat{\bm\rho}(B,\bm\pi)\big)$.
Because different values of $B$ lead to different numbers of parameters in the model,
we cannot estimate the parameters by maximizing (\ref{eq:llike}).  Instead we view this as a model selection problem and again use the MDL principle. In this way we can choose the ``best'' combination of $B$ and $\bm{\pi}$, and then compute $\hat\gamma_b(B, \bm\pi)$, $\hat\rho_b(B, \bm\pi)$, and $\hat\boldtheta_b(\hat\gamma_b, \hat\rho_b)$ \changes{for each time interval}. We derive the MDL criterion  in Appendix~\ref{app:mdl2} and show that the best model can be found by minimizing
\begin{align}
  \mathsf{mdl}_{\rm full}(B, \bm{\pi})
&= -\ppplike(B, \bm{\pi}) + \log {B} + \sum^{{B}}_{b=1}\log{c_b(\bm\pi)} \nonumber\\
&\quad\quad+
\sum^{{B}}_{b=1}\left[\frac{1}{2}\| \hat\boldtheta_b(B,\bm\pi)\|_0\log\{Nc_b(\bm\pi)\} + \log {N \choose \|\hat{\bm{\eta}}_b(B,\bm\pi)\|_0}\right], \label{eqn:mdl_break}
\end{align}
where ${c}_b(\bm\pi)$ is the cardinality of $\{ i: {\pi}_{b-1}\le t_i< {\pi}_{b}\}$ for $b=1,\dots,{B}$, i.e.,
${c_b(\bm\pi)}$ is the number of time bins in time interval $b$.

\subsection{Practical minimization}
\label{sec:fullfit}
We now consider  minimization of the MDL criterion given in~(\ref{eqn:mdl_break}).  
Once the number of time intervals, $B$, and the locations of change points, $\bm{\pi}$, are specified, unique estimates of $\bm{\gamma}$ and $\bm{\rho}$ can be obtained using the method described in Section~\ref{sec:hom}.  
Since the time intervals are independent the required computation is highly parallelizable.  Nonetheless, minimization of~(\ref{eqn:mdl_break}) is not trivial. It involves an expensive combinatoric optimization: for each $B$, there are ${J-1}\choose{B-1}$ possibilities for $\bm{\pi}$. (A small number of these possibilities is not considered by our optimiziation procedure because we impose a minimum width on each time interval.)
To simplify computation, we propose a fast stepwise forward algorithm to approximately minimize~(\ref{eqn:mdl_break}), rather than attempting to globally minimize it.  Briefly, at each step we add a single change point to the current collection by selecting the new change point that decreases $\mathsf{mdl}_{\rm full}(B,\bm{\pi})$ the most. 

\begin{algorithm}[t]
  \caption{A stepwise search algorithm for minimizing~(\ref{eqn:mdl_break})}\label{alg:treegrow}
  \begin{flushleft}
  \textbf{Input:} Wavelength-cross-time bin counts $\{(\Cnt)\}_{i=1,\dots,N, j=1,\dots, J}$\\
  \textbf{Description:} A fast algorithm for approximately minimizing $\mathsf{mdl}_{\rm full} (B, \bm{\pi}, \bm{\gamma}, \bm{\rho})$
\end{flushleft}
  \begin{algorithmic}[1]
    \State $B \leftarrow 1$ and $\bm{\pi} \leftarrow ()$ (null vector)
        \State $z \leftarrow {\mathsf{mdl}}_{\rm full} (1, \bm{\pi})$.
          \State $z' \leftarrow z + 1$
        \While  {$z < z'$}
        \State $B \leftarrow B+1$
        \State $z' \leftarrow z$
        \State $\bm{\pi}' \leftarrow \bm{\pi}$
        \State Use a grid search to find the change point that when added to the current collection, $\bm{\pi}'$, gives the largest reduction in ${\mathsf{mdl}}_{\rm full}$. Denote this new set of change points as $\bm{\pi}$.
        If there is no possible change point to add, break the while-loop.
        \State $z \leftarrow {\mathsf{mdl}}_{\rm full}(B, \bm{\pi})$
        \EndWhile
        \State {\bf end while}
    \State {Return $\{B-1, \bm{\pi}'\}$}
\end{algorithmic}
\end{algorithm}

More precisely, the algorithm first calculates (\ref{eqn:mdl_break}) for the model with no change points (i.e., only one time interval), that is, it computes $\mathsf{mdl}_1 =  \mathsf{mdl}_{\rm full}(1,\bm{\pi})$. (This does not depend on $\bm\pi$ because there are no change points.)  In the second step, the algorithm considers all of the possible models with a single change point and selects the one that minimizes $\mathsf{mdl}_{\rm full}(2,\bm{\pi})$.
Calling the obtained  minimum $\mathsf{mdl}_2$, the change point is added to the model if $\mathsf{mdl}_2 < \mathsf{mdl}_1$; otherwise the algorithm terminates.  The algorithm continues in this way, attempting to add an additional change point to the model at each step by minimizing (\ref{eqn:mdl_break}). It stops when there is no possible change point remaining that decreases $\mathsf{mdl}_{\rm full}(B,\bm{\pi})$ and this current model is taken as the (approximate) minimizer of~(\ref{eqn:mdl_break}).  Although this greedy algorithm may result in local minimization,  it is fast and provides a good compromise between computational efficiency and accuracy \citep{Lee02:genetic}.  The exact steps of the algorithm appear  in Algorithm~\ref{alg:treegrow}.

\subsection{A Monte Carlo Procedure for Testing for the Existence of Change Points}
\label{sec:MCtest}

\begin{algorithm}[t]
  \caption{Monte Carlo significance test}\label{alg:mctest}
  \begin{flushleft}
  {\textbf{Input:} Working data $\{(w_i, t_j, Y_{ij})\}_{i=1,\dots,N, j=1,\dots, J}$, number of simulation $N_{\mathrm{sim}}$, significance level $\alpha$}\\
  {\textbf{Description:} Monte Carlo test for change point}
\end{flushleft}
  \begin{algorithmic}[1]
  \State Fit the time homogenous Poisson moded (with no change point, Section \ref{sec:hom}) and full Poisson model (including change points, Section~\ref{sec:fullfit}) to $\{(w_i, t_j, Y_{ij})\}_{i=1,\dots,N, j=1,\dots,
    J}$ and denote the test statistic
  $m^\star = \mathsf{mdl}_{\rm full}(\hat B, \bm{\hat\pi})-\mathsf{mdl}_{\rm
  full}(1, \bm{\pi}_{\rm null})$. ($\bm{\pi}_{\rm null} = ()$, null vector)
    \For {$k=1$ to $N_{\mathrm{sim}}$}
    \State Generate the data set $\mathcal D_k = \{(w_i, t_j, Y_{i p_j})\}$ where $\{p_1,\dots,p_J\}$ is a uniformly random permutation of $\{1,2,\dots, J\}$.
      \State Fit the full Poisson model (including change points) using the procedure in Section~\ref{sec:fullfit} to  $\mathcal D_k$.
      \State Denote the resulting test statistic by $m^\star_k$.
    \EndFor
  \State Estimate the $p$-value as
    \[
      \hat{p} = \frac{\#\{m_k^\star \ge m^\star\} + 1}{N_{\mathrm{sim}}+1}
    \]
  \State Conclude that there is at least one change point if $\hat{p} < \alpha$ for some significance level, $\alpha$.
\end{algorithmic}
\end{algorithm}

In practice it is useful to quantify evidence for existence of the change points. 
For a test statistic, we use the change in the
MDL function (\ref{eqn:mdl_break})
due to introduction of change points, 
$m^\star = \mathsf{mdl}_{\rm full}(\hat B, \bm{\hat\pi})-\mathsf{mdl}_{\rm
  full}(1, \bm{\pi}_{\rm null})$,
where
$\hat B$ and $\bm{\hat\pi}$ are computed by minimizing (\ref{eqn:mdl_break}),
and $\bm{\pi}_{\rm null}$ is the null vector containing no change point. 
Here $\mathsf{mdl}_{\rm full}(1,\bm{\pi}_{\rm null})$ is the value of the MDL
function (\ref{eqn:mdl_break}) with no change points.
We approximate the null distribution of $m^\star$ via Monte Carlo using permutation.
Specifically, we generate $N_{\rm sim}$ uniformly random permutations of $\{1,2,\dots, J\}$
independently and the corresponding replicate null
datasets, $(\mathcal D_1, \ldots, \mathcal D_{N_{\rm sim}})$,
by shuffling the time indices of the original dataset according to these permutations (see Step 3 of Algorithm~\ref{alg:mctest}).
Each replicate dataset is then fit with the full model (including change
points) according to the MDL criterion specified in (\ref{eqn:mdl_break}) and
the resulting test statistic, $m^\star_i$ is computed (for $i=1,\ldots, N_{\rm
  sim})$.
Note that the MDL criterion (\ref{eqn:mdl_break}) with no change points,
$\mathsf{mdl}_{\rm full}(1,\bm{\pi}_{\rm null})$,
is invariant to the permutation and thus need not to be re-computed for the permuted datasets
$(\mathcal D_1, \ldots, \mathcal D_{N_{\rm sim}})$.
The empirical distribution of $(m^\star_1, \ldots, m^\star_{N_{\rm
    sim}})$ is a Monte Carlo approximation to the null distribution of
$m^\star$ and a p-value can be computed by comparing $m^\star$ to $(m^\star_1,
\ldots, m^\star_{N_{\rm sim}})$. Details appear in Algorithm~\ref{alg:mctest}.

\section{Numerical Experiments}
\label{sec:sim}
We conducted a series of simulation experiments  to evaluate the empirical properties of the proposed methodology.  
Binned photon counts were simulated under \changes{eight} different spectral-temporal test functions, $\lam$.  In order to make the experiments as realistic as possible, \changes{seven of the test functions} were constructed by fitting a spectral-temporal model, using the method described in Section \ref{sec:break}, to the seven observations of \fkcom\  described in Section~\ref{sec:real}.
\changes{Because these test functions have at most two change points and because we would like to investigate the statistical properties of our method with more change points, an eighth test function was created by concatenating two of the original test functions. (I particular, those created from
observation identification (ObsID) numbers 13251 and 12298, see Table~\ref{tab:tfdetails}.)}
\changes{The analyses carried out here, including the binning of the data, is identical to that described in Section~\ref{sec:real}.}
\changes{Further details are summarized in Table~\ref{tab:tfdetails}.}
Together \changes{these eight test functions} cover a wide range of possible scenarios, ranging from models with no change point to models with \changes{four} change points.  For each of the \changes{eight} $\lam$, 200 data sets were simulated according to~(\ref{eq:pois-mdl}), using the real data values for $\delta_t$, $\delta_w$, and $A(w_i)=\sum^K_{k=1}A_k(\wl)$.  The proposed methodology was applied to all $\changes{8}\times 200 =1400$ simulated data sets  to estimate $\lam$ for each.

\begin{table}[t]
\begin{center}
  \changes{\caption{Details of simulation settings.
The columns report (1) the
chronologically ordered (except the last row)
ObsID numbers of the
datasets used to generate each test function,
(2) the number of wavelength bins,
(3) the number of time bins,
(4) the number of basis functions $P$, as in Equation (\ref{eqn:gensemi}),
used to fit the corresponding FK Com observation,
(5) the number of time intervals, i.e., the number of change points plus one,
estimated from the corresponding FK Com observation,
(6) vector of the numbers of emission lines with element $j$ corresponding
to time interval $j$,
estimated from the corresponding FK Com observation.
}
\label{tab:tfdetails}
\begin{tabular}{cccccc}
  \hline
ObsID & $N$ & $J$ & $P$ & $B$ & $M$ \\
  \hline
 12297 & 142 & 21 & 34 &  2 & (0, 1)  \\
 12356 & 142 & 32 & 34  & 1  & (1)\\
 13251 & 142 & 49 & 34  & 3 & (0, 0, 0)\\
 12298 & 142 & 20 & 34  & 2 & (2, 1)\\
 13259 & 142 & 23 & 34  & 2 & (0, 0)\\
 12357 & 142 & 18 & 34 & 1  & (1)\\
 12299 & 142 & 20 & 34  & 3 & (0, 0, 0)\\
 13251, 12298 & 142 & 69 & 34  & 5 & (0, 0, 0, 2, 1)\\
   \hline
\end{tabular}}
\end{center}
\end{table}

The simulation results are summarized in Table~\ref{tab:sim}.  The first column lists \changes{the ObsID number} for the observations of \fkcom\ (in chronological rather than ascending order) used to generate the $\lam$, see Section~\ref{sec:real}.  The second column shows the (true) numbers of homogeneous time intervals, $B$, in each of the $\lam$; recall the number of change points is $B-1$. The third column reports the percentage of the simulated datasets for which the fitted $\hat B$ exactly equals the true value. The proposed methodology achieved a correct recovery rate of at least 94\% in 6 out of the \changes{8} simulations settings.

\begin{table}[t]
\begin{center}
\caption{Simulation results. The columns report (1) the chronologically ordered \changes{(except the last row)} ObsID numbers
 of the datasets 
used to generate each test function, 
(2) the true number of time intervals, i.e., the true number of change points plus one,
(3) the percentage of the fitted $\hat{B}$ that exactly equal the true $B$, and
(4)  the root mean square error of $\hat{\bm{\pi}}$ among those simulations with $\hat{B}=B$.}
\label{tab:sim}
\begin{tabular}{cccccc}
  \hline
ObsID & $B$ & \% correct $\hat{B}$ & RMSE($\hat{\bm{\pi}}$)\\
  \hline
 12297 & 2 & 94 & 0.70\\
 12356 & 1 & 100 & 0\\
 13251 & 3 & 98 & 0.66\\
 12298 & 2 & 98 & 0.29\\
 13259 & 2 & 80 & 0.69\\
 12357 & 1 & 100 & 0\\
 12299 & 3 & 100 & 0\\
 \changes{13251, 12298} & \changes{5} & \changes{85} & \changes{0.66}\\
   \hline
\end{tabular}
\end{center}
\end{table}

Our method also accurately estimates the change points, $\bm{\pi}$, \changes{as shown in Column~4 of Table~\ref{tab:sim}. For those simulations where $B$ is correctly estimated,
Column~4 gives the root mean squared error (RMSE) of $\hat{\bm{\pi}}$, in terms of the time bin width ($\delta_t =2000$ seconds in all cases).}   The RMSE is less than one bin width in all \changes{eight} cases.  Figure~\ref{fig:sim} displays histograms of the estimated change point location  among those simulations for which $\hat B=B$ under the \changes{five} test functions with less than a 100\% recovery rate  for $\bm{\pi}$. In  all \changes{five} cases, the estimated change point locations are narrowly centered around the true change points.
Finally, for those simulations for which $\hat{{B}}={B}$ and $\hat{\bm{\pi}}=\bm{\pi}$,
the uncertainty of $\hat{f}$ is summarized 
in Figure~\ref{fig:cont}.  
Here we only present results for one test function (ObsID~12299); plots for the other test functions appear in 
Appendix~\ref{app:sim}.  For each of the $B=3$ homogeneous time intervals associated with ObsID~12299, 
we overlay the fitted  $\hat{f}$ of the simulations  for which 
$\hat{{B}}={B}$ and $\hat{\bm{\pi}}=\bm{\pi}$. The results are plotted in panels (a)-(c) of Figure \ref{fig:cont}. The relatively high uncertainty at high wavelengths is partly due to the small effective area at these wavelengths; see panel (d) of Figure \ref{fig:cont}.

\begin{figure}[t]
  \centering
\includegraphics[width=0.8\linewidth]{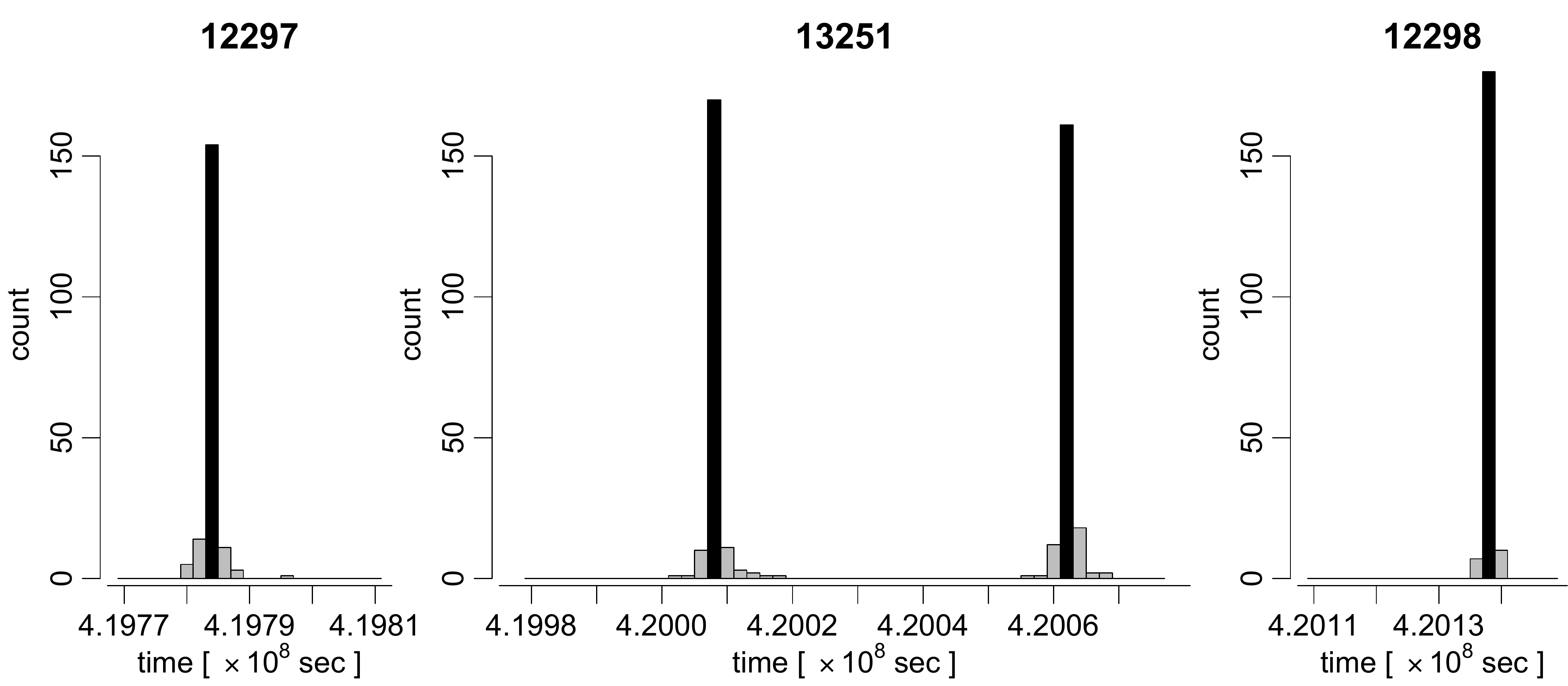}
\includegraphics[width=0.8\linewidth]{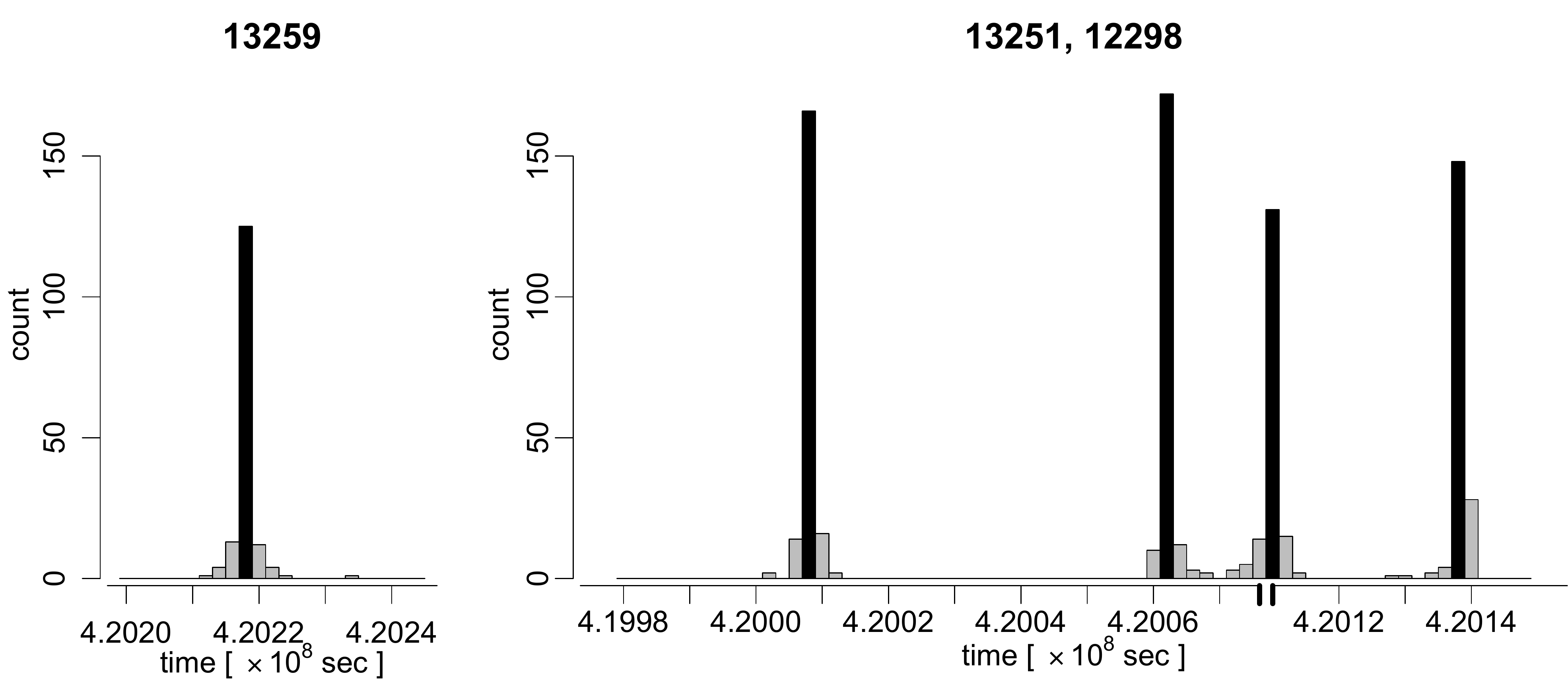}
\caption{Histograms of detected change point locations in the simulation study.
  The title of each panel refers to the ObsIDs  \changes{given in Tables~\ref{tab:tfdetails} and \ref{tab:sim}. The widths of the histogram bins is the same as the time binning of the data; 
  bins containing} the true change-points locations are represented in black.
  The horizontal axis \changes{is elapsed spacecraft time and the panel corresponding to the concatenation of ObsID~13251 and ObsID~12298 has a broken x-axis because there is a gap in spacecraft time between the two observations.}}\label{fig:sim}
\end{figure}

\begin{figure}[p]
  \centering
\includegraphics[width=0.49\linewidth]{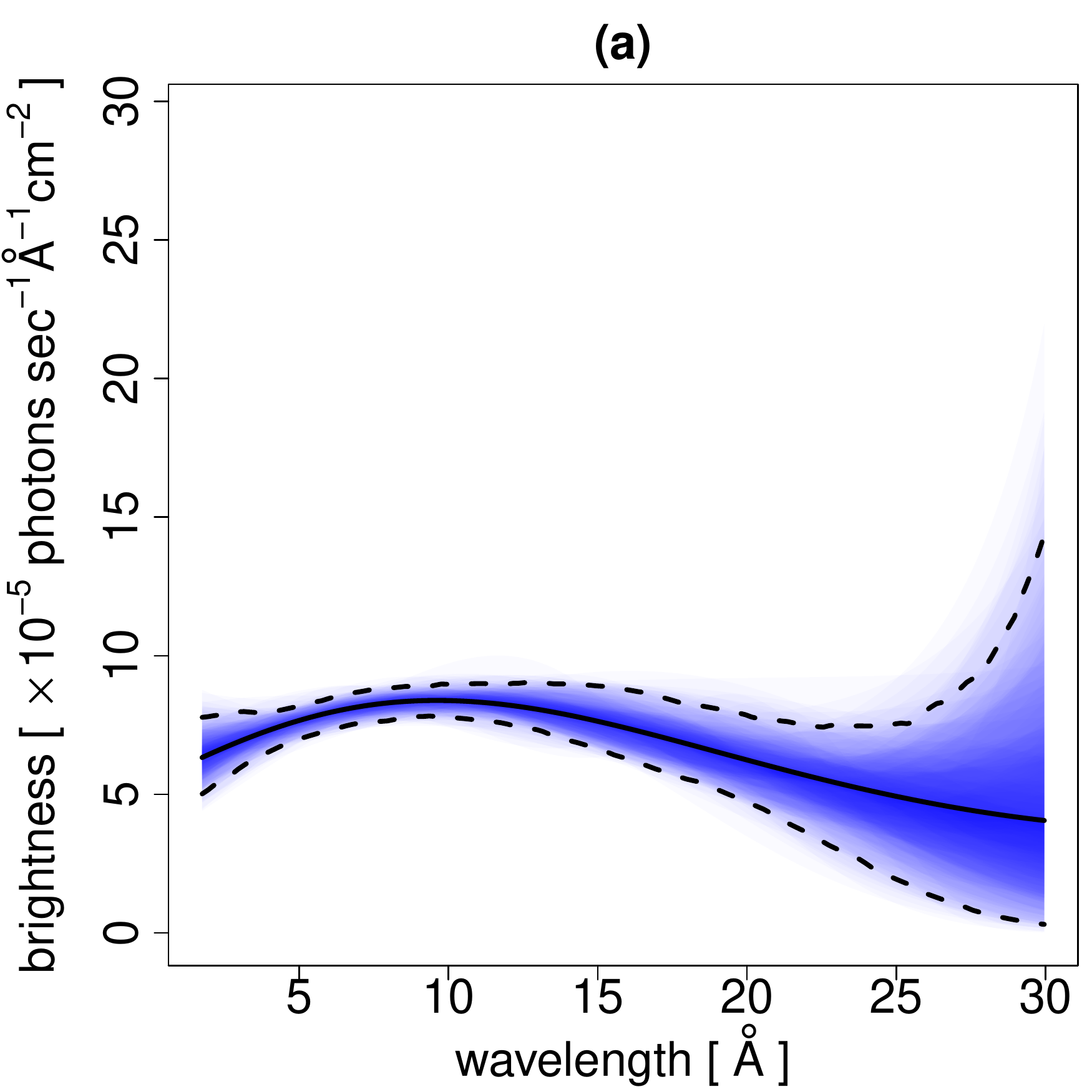}
\includegraphics[width=0.49\linewidth]{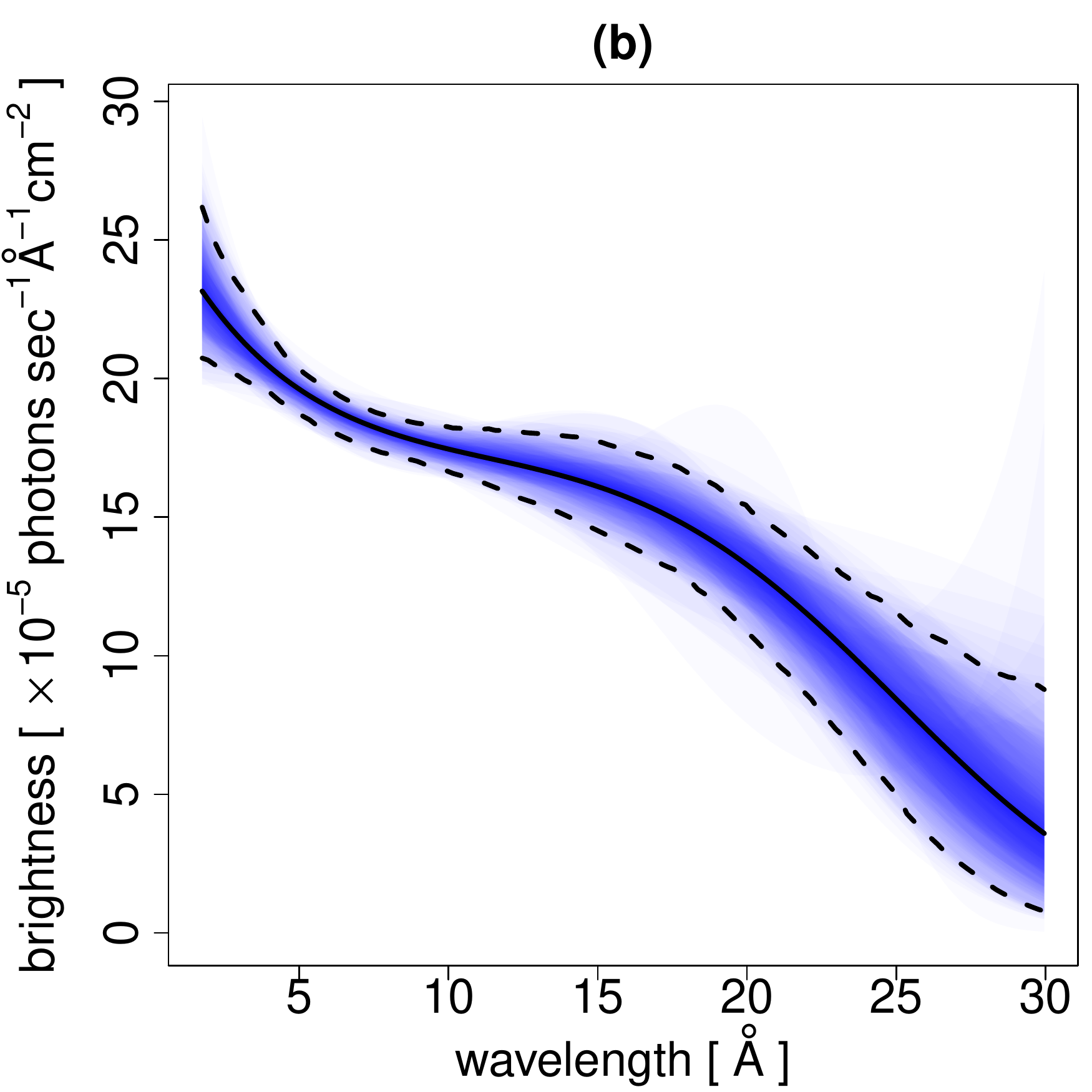}\\
\includegraphics[width=0.49\linewidth]{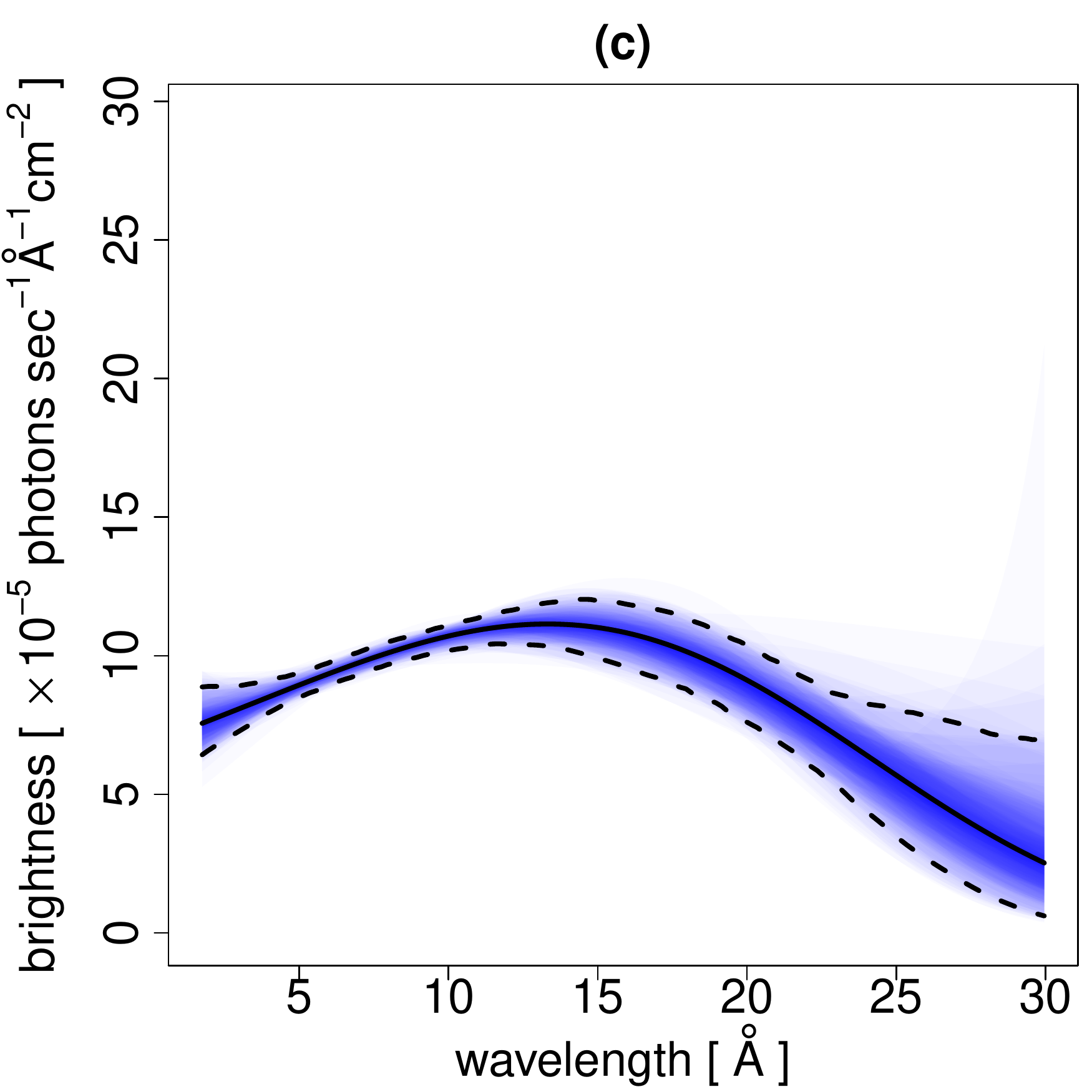}
\includegraphics[width=0.49\linewidth]{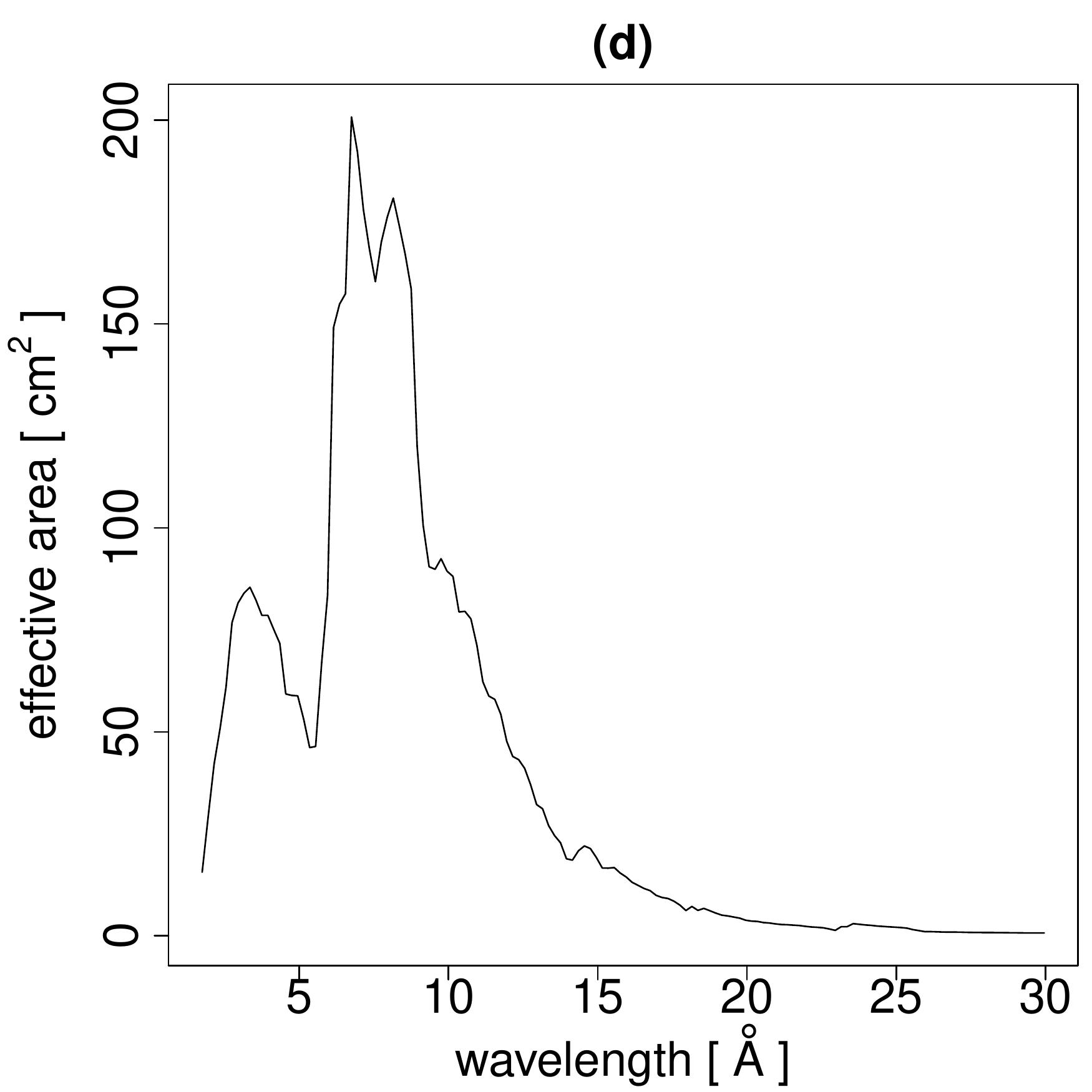}
\caption{
  \changes{Panels (a) - (c):  Overlay plots of the $\hat{f}$ fitted to the 200 simulations generated with the test function corresponding to ObsID~12299. 
  Notice that all 200 simulations resulted in $\hat{B}=B$ and $\hat{\bm{\pi}}=\bm{\pi}$.
  The three panels represent the $B=3$ homogeneous time intervals of ObsID~12299.
  The areas between the $\hat{f}$ computed with each
  simulated dataset and the underlying test function are colored and overlaid.
  Darker color represents more overlap of these areas  and thus summarizes the
  concentration of the 
  the fitted spectra  around the underlying test function. 
  The solid black line represents the
  smooth component, $f$, of test function 12299 and the dashed lines represent
  the 5\% and 95\% (pointwise) percentiles of $\hat{f}$.
  Similar plots for the other test functions appear in Appendix~\ref{app:sim}.
  Panel (d): Plot of the effective area $\sum^K_{k=1} A_k(w)$ curve for test function 12299.}
}\label{fig:cont}
\end{figure}

\section{Application to X-ray Observations of \fkcom}
\label{sec:real}

\fkcom\ is an evolved yellow giant star (spectral type G4\,{\tt III}) with a
mass of about  $8.4$ times that of the sun and at a distance of
$550\pm60$~light-years. Typically, late-spectral-type giants are slow rotators,
and thus do not maintain a strong magnetic dynamo that can sustain a
high-temperature corona. However, \fkcom\ is an extremely fast rotator, with
equatorial velocities of 179~km/s, and a rotational period of 2.4~days
\citep{korh:etal:99,stras:09}.  (For comparison, the Sun rotates at about 2 km/s at its equator.)  Giant stars are formed when a sun-like star
depletes the hydrogen at its core and thus transitions from producing energy
though the fusion of hydrogen into helium to the fusion of heavier elements.
Typical sun-like progenitor stars would require rotational speeds that would
tear them apart in order to form a giant star with such a rapid rate of spin.
Thus, the favored hypothesis for the formation of \fkcom-like stars is that they are the result of a binary star system that has coalesced through a process of mass transfer, dramatically speeding up the stellar rotation  in the process \citep{bopp:sten:81}.
If this is the case, the high rotation should induce a strong dynamo and consequently a dynamic corona, filled with high-temperature plasma and organized into loop-like structures via magnetic fields. 

\fkcom\ exhibits long duration spots---akin to sunspots---and flares associated with these spots. Optical Doppler imaging observations of \fkcom\ demonstrate the existence of persistent spot-like features \citep{Elstner-Korhonen05} that remain localized to specific longitudes for long timescales ($>$months).
Analysis of an X-ray flare observed with the XMM-{\sl Newton} observatory showed that the flare is tied to these spots \citep{Drake-Chung-Kashyap08}.
Numerical MHD modeling of the stellar wind and the magnetic structure \citep{Cohen-Drake-Kashyap10} shows a highly complex coronal environment, with the most notable feature being even putatively stable coronal loops ``wrapping around'' due to the rotation.
Thus, even though the corona appears to be solar-like in its origin, it is an extreme case of such coronae, and is therefore of considerable interest to both theoretical and observational studies.

A long-duration, high-resolution X-ray observation of \fkcom\ was carried out with the {\sl Chandra X-ray Observatory}  \changes{over seven discontinuous time periods} in April 2011.
The data show that the source brightness is constantly fluctuating with occasional periods of singular flaring activity, see Figures~\ref{fig:fkcom_lc} and \ref{fig:data}.
As a test case for our method,  \changes{we apply it to these seven {\sl Chandra} observations of \fkcom.} Each was preprocessed by discarding photons with
 wavelength less than 1.65  or greater than 31~\ang.
\changes{The natural resolution of {\sl Chandra}  data of this sort is extremely high, resulting in a very large, but very sparse table of photon counts; typically there is at most one photon recorded in each wavelength-cross-time bin. We can obtain substantial computational savings with only a modest sacrifice of scientific information by reducing the resolution of the data. Massive events in the corona of \fkcom, for example, require appreciable time and are associated with dramatic spectral shifts. Change points typically appear with frequencies on the order of tens of thousands of seconds, whereas the inherent bin width is about 3 seconds. In our analyses, we reduce this resolution and use temporal bins of width $\delta_t=2000$~seconds. Similarly, the inherent wavelength bin width is 0.005~\ang\ and we set $\delta_w=0.2$~\ang. Though a substantial reduction in spectral resolution, our analysis is much higher resolution than the standard hardness ratio analysis, which can be viewed as using two or three wavelength bins of width 10-15~\ang.}
Our fitted $\lambda(t_j,w_i)$ are plotted in Figure~\ref{fkcoms2}.

\begin{figure}[t]
\begin{center}
\includegraphics[width=0.8\linewidth]{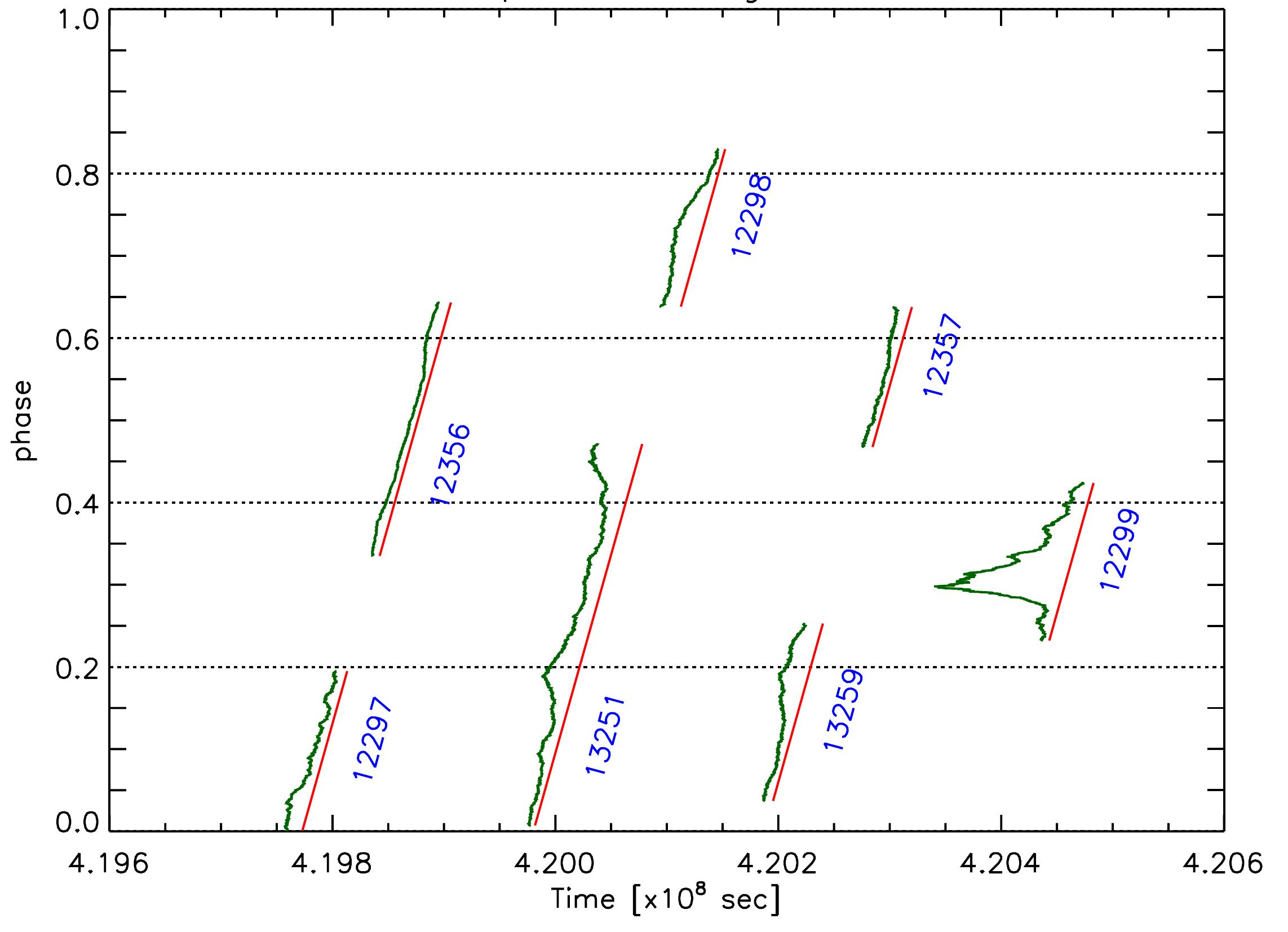}
\caption{The phase-time light curve of \fkcom.
The source brightness (i.e., light curve) over the course of the {\sl Chandra} observation is shown during periods when the source was observed.
The horizontal axis represents spacecraft elapsed time, and the vertical axis represents the rotational phase of the star.
The tilted red lines show the phase-time correspondence for each observation, showing the time and rotational phase coverage of the full dataset. 
Overlaid on the phase-time line, displayed as a deviation from it, are sparklines depicting the   smoothed light curves (green curves).
The segments are labeled by the ObsID numbers.
This representation of the light curve shows that there is no obvious rotational signature in the light curve, and the variability is stochastic.
Notice the large flare that occurs in ObsID~12299: our analysis shows that the spectrum varies significantly during this flare, see Figure~\ref{fkcoms2}.
}
\label{fig:fkcom_lc}
\end{center}
\end{figure}

\begin{table}[H]
\begin{center}
\caption{Monte Carlo p-values for the test for change points in each of the seven observations of \fkcom}
\label{tab:pvalue}
\begin{tabular}{cccccccc}
  \hline
ObsID & 12297 & 12356 & 13251 & 12298 & 13259 & 12357 & 12299\\
p-value & 0.12 & 1.00 & $<$0.01 & 0.01 & 0.14 & 1.00 & $<$0.01\\
   \hline
\end{tabular}
\end{center}

\end{table}

We also applied the Monte Carlo test for change points as described in Section~\ref{sec:MCtest}; the resulting p-values for the seven datasets appear in Table~\ref{tab:pvalue}.
There is no evidence for change points in two of the observations (ObsID~12356
and 12357), very weak evidence for two (ObsID~12297 and 13259), and strong
evidence for the other three. This agrees with the fitted models illustrated in
Figure~\ref{fkcoms2}, but quantifies the degree of evidence for the fitted
change points.  Finally, the two test functions that resulted in relatively low
recovery rates of $B$ in the simulations described in Section \ref{sec:sim}
correspond to the two data sets with the weakest evidence for the a change
point, compare ObsID 12297 and 13259 in Table~\ref{tab:sim} (\% correct $\hat{B}$)
with Table~\ref{tab:pvalue} (p-value).

\begin{figure}[t]
\begin{center}
\includegraphics[width=\linewidth]{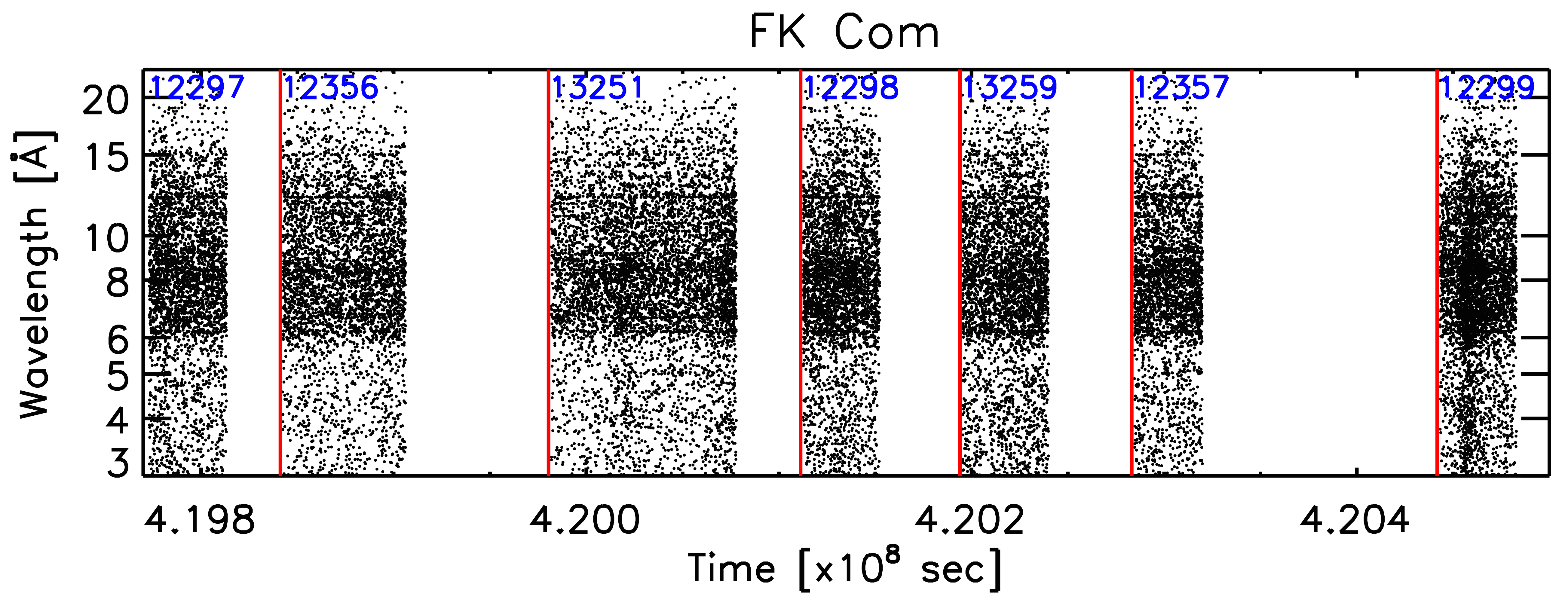}
\caption{\changes{Observed data. Each point represents a photon detected with HETGS+ACIS-S configuration of {\it Chandra} (including both low- and high-energy
gratings data).  The wavelengths are given in \ang, and the time
is in spacecraft clock seconds.  The ObsIDs of each segment
are marked to the right of the red vertical line representing the
starting time of the observation segment.  The collected data are sparse, but some
changes in the density of points can be discerned even by eye,
indicating temporal and spectral variations.}}
\label{fig:data}
\end{center}
\end{figure}

\changes{Although we find two or fewer change points in each observation, the number
was not limited in our analyses. 
Operational constraints limit the continuous observation time available with 
the space-based {\it Chandra X-ray Observatory}. 
The change points that we observe stem from massive energetic shifts in the cornea of \fkcom\ and take time to evolve. Keeping in mind that a single massive shift of this sort on the Sun would likely destroy human civilization, observing a total of seven in one (discontinuous) observation period spanning less than a fortnight (see Figure~\ref{fig:fkcom_lc}) is not a small number.
}

\begin{figure}[p!]
  \centering
\includegraphics[width=0.9\linewidth]{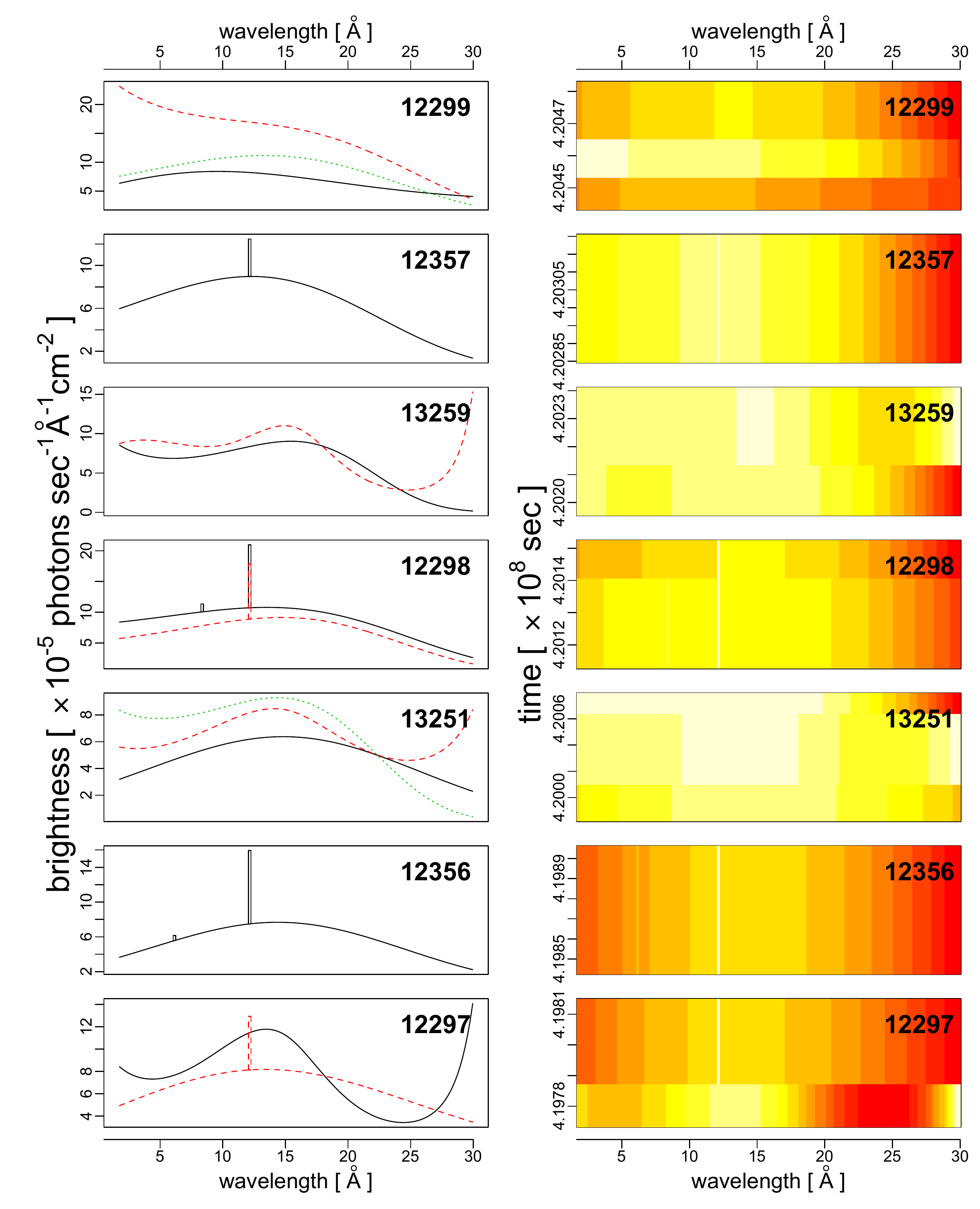}
\caption{Fitted $\lam$ for seven {\sl Chandra} observations of \fkcom. The ObsID of each dataset appears in the title of each panel. The left plot in each row shows  the fitted spectra for each identified time interval; the solid black, dashed red, and dotted green curves correspond to the first, second, and third time intervals, respectively.    The right plot in each row is a heat maps that represents the best fit values of $\lam$. Time is given in elapsed spacecraft time. 
\label{fkcoms2}
}
\end{figure}

This analysis indicates that large changes in intensities are accompanied by significant changes in the spectrum of the source.  This is readily apparent in the results for ObsID 12299, which shows the observation separated into three segments with different intensities and spectral shapes; see ObsID~12299 in Figure~\ref{fig:fkcom_lc} and row 7 of Figure~\ref{fkcoms2}. The spectrum changes to being dominated by high-energy photons due to an increase in high-temperature plasma when the flare is set off.  As the flare decays, even when the observed intensity is relatively high, the spectrum can be seen returning to the pre-flare state, though with enhanced emission over the line-dominated region around $15$~\ang.

From a theoretical point of view, this is not an obvious result, since for active stars like \fkcom, it is believed that even the apparently quiescent corona emits X-rays as a superposition of a large number of weak flares \citep[see, e.g.,][]{Kashyap-Drake-Gudel02, Gudel04}.
However, our analysis favors the idea that hot plasma could become trapped in stretched field lines wrapping around the star, and thus cool radiatively, and flares are only triggered sporadically when the magnetic strain becomes too large to be sustained.
While a detailed modeling and interpretation of this picture is beyond the scope of this paper,  we point out that even a simple application of a statistically principled method is enough to obtain an important result: The quiescent corona of \fkcom\ is {\sl unlike} the flaring corona, and in this sense is much more solar-like than previously anticipated.

\section{Discussion}

\label{sec:conc}
This article develops a novel approach for modeling change points in a class of marked Poisson processes, specifically the time-varying spectra of high-energy astrophysical sources.  The approach includes a family of change point models, a practical algorithm for estimation, and an MDL criterion for selecting the best fitting model. Despite numerous reported applications where the MDL principle leads to reliable model selection criteria, its use in the statistical literature remains rather sparse.  It is our hope that the present work, in which the MDL principle is applied to solve a complicated model selection problem that involves the ``large $p$ small $n$'' scenario, will encourage and stimulate renewed research interests in this useful principle among statisticians. 

Although the methods developed in this article are motivated by the search for change points in high-energy time-varying spectra, it is possible to extend them to include spatial variations as well. Typically, astrophysical images include sources at many different scales, ranging from unresolved point sources to complex structures that may span the field of view. Owing to instrumental point-spread functions, point sources in close proximity may overlap and in many cases are overlaid on larger scale structures.  An extension 
that can account for change points in the spatial plane or in the spatial cross spectral hyperplane would be quite useful in astronomical data analysis.  Statistically, this involves extending the dimension of the marks in the Poisson process from the one dimensional photon wavelengths to two dimension spatial locations or three dimensional spatial cross spectral measurements. This would, of course, involve extensions of the non-parametric models developed here for spectra to higher dimensional models for images or joint models for images and spectra.

\subsubsection*{Acknowledgments}

\changes{The authors are grateful to the Referees, the Associate Editor and the Editor, Professor Nicoleta Serban, for their many useful and constructive comments. They substantially improved the paper. This work was conducted under the auspices of the CHASC International Astrostatistics Center.  CHASC is supported by NSF grants DMS-12-08791, DMS-12-09232 and DMS-1513484.  In addition, Vinay Kashyap acknowledges support from Chandra grant GO1-12021X, and from NASA's contract to the Chandra X-ray Center, NAS8-03060, Thomas Lee from National Science Foundation grants DMS 12-09226 and DMS 15-12945, David van Dyk from a Wolfson Research Merit Award provided by the British Royal Society and from a Marie-Curie Career Integration Grant provided by the European Commission.
We also thank Xiao-Li Meng, Tom Ayres, Ofer Cohen, Jeremy Drake,
David Garcia-Alvarez, Dave Huenemoerder, Heidi Korhonen, Steve Saar,
Paola Testa, and Aad van Ballegooijen for many useful discussions.}

\bigskip
\appendix
\section{Derivation of the MDL criterion~(\ref{eqn:mdl_hom})}
\label{app:mdl1}
This appendix derives (\ref{eqn:mdl_hom}), and it begins with the calculation of the term $\mathsf{CL}(\hat{\mathcal{M}})$ in~(\ref{eqn:cl}).  In Section~\ref{sec:hom}, every candidate model $\mathcal{M}$ can be uniquely identified by a $\bm{\theta}$ and thus $\mathsf{CL} (\hat{\mathcal{M}}) = \mathsf{CL}(\hat{\bm{\theta}})$.  Now to completely describe an $\hat{\bm{\theta}}=(\hat{\bm{\beta}}^\tp, \hat{\bm{\eta}}^\tp)^\tp$, one could first specify which elements are nonzero, and then specify the actual values of these nonzero elements.  As there are ${P \choose \|\hat{\bm{\beta}}\|_0}$ ways of arranging $\|\hat{\bm{\beta}}\|_0$ nonzero $\hat{\beta}_j$'s in $P$ ``slots'', we need $\log_2 {P \choose \|\hat{\bm{\beta}}\|_0}$ bits to specify which $\hat{\beta}_j$'s are nonzero.  Similarly we need $\log_2 {N \choose \|\hat{\bm{\eta}}\|_0}$ bits to specify which $\hat{\eta}_j$'s are nonzero.

Now for the actual values of the nonzero elements in $\hat{\bm{\theta}}$.  As they are penalized maximum likelihood estimates, the approximate effective codelength for each of them is $\frac{1}{2} \log_2 (NJ)$ \citep[e.g.,][]{Rissanen89}, giving the total codelength for all nonzero elements as $\frac{1}{2} \|\hat{\bm{\theta}}\|_0 \log_2 (NJ)$.  Therefore 
\[
  \mathsf{CL}(\hat{\mathcal{M}})= \mathsf{CL}(\hat{\bm{\theta}}) =
\frac{1}{2}\|\hat{\bm{\theta}}\|_0\log_2 (NJ) +
  \log_2 {P \choose \|\hat{\bm{\beta}}\|_0} +
  \log_2 {N \choose \|\hat{\bm{\eta}}\|_0}.
\]
Since in practice $N \gg P$ so the second term is ignored, giving our final expression for $\mathsf{CL}(\hat{\mathcal{M}})$:
\[
  \mathsf{CL}(\hat{\mathcal{M}})= \frac{1}{2}\|\hat{\bm{\theta}}\|_0\log_2 (NJ) +
  \log_2 {N \choose \|\hat{\bm{\eta}}\|_0}.
\]
We remark that in many classical applications of two-part MDL where the number of possible parameters is small when compared to the number of data points, a penalty term similar to the last term in the above expression is typically ignored.  However, for the present problem the number of potential parameters $P+N$ is not ignorable when compared to the number of data points $NJ$ (where $J$ could be small due to change points, see Section \ref{sec:break}). Failing to consider this last term will lead to an overfitted model.  This last term shares the same role as the additional penalty term in the Extended Bayesian Information Criterion (EBIC) proposed by \citet{Chen-Chen08}.  These authors have shown that the traditional BIC fails in the so-called ``large $p$ small $n$'' scenario and an additional penalty term is required to guarantee consistency properties.

Lastly we need to calculate the term $\mathsf{CL}(\mathcal{D}|\hat{\mathcal{M}})$, which has been shown by \citet{Rissanen89} that it is given by the negative of the log likelihood.  In our case this gives $\mathsf{CL}(\mathcal{D}|\hat{\mathcal{M}})= -\sum^N_{i=1} \sum^J_{j=1}l_{\mathrm{one}}(Y_{ij};\hat{\bm{\theta}})$ with base 2 logarithm.  Combining the expressions for $\mathsf{CL}(\hat{\mathcal{M}})$ and $\mathsf{CL}({\mathcal{D|\hat{\mathcal{M}}}})$ and changing the base of the logarithm from 2 to $e$, we obtain~(\ref{eqn:mdl_hom}).

\begin{figure}[H]
  \centering
\includegraphics[width=0.25\linewidth]{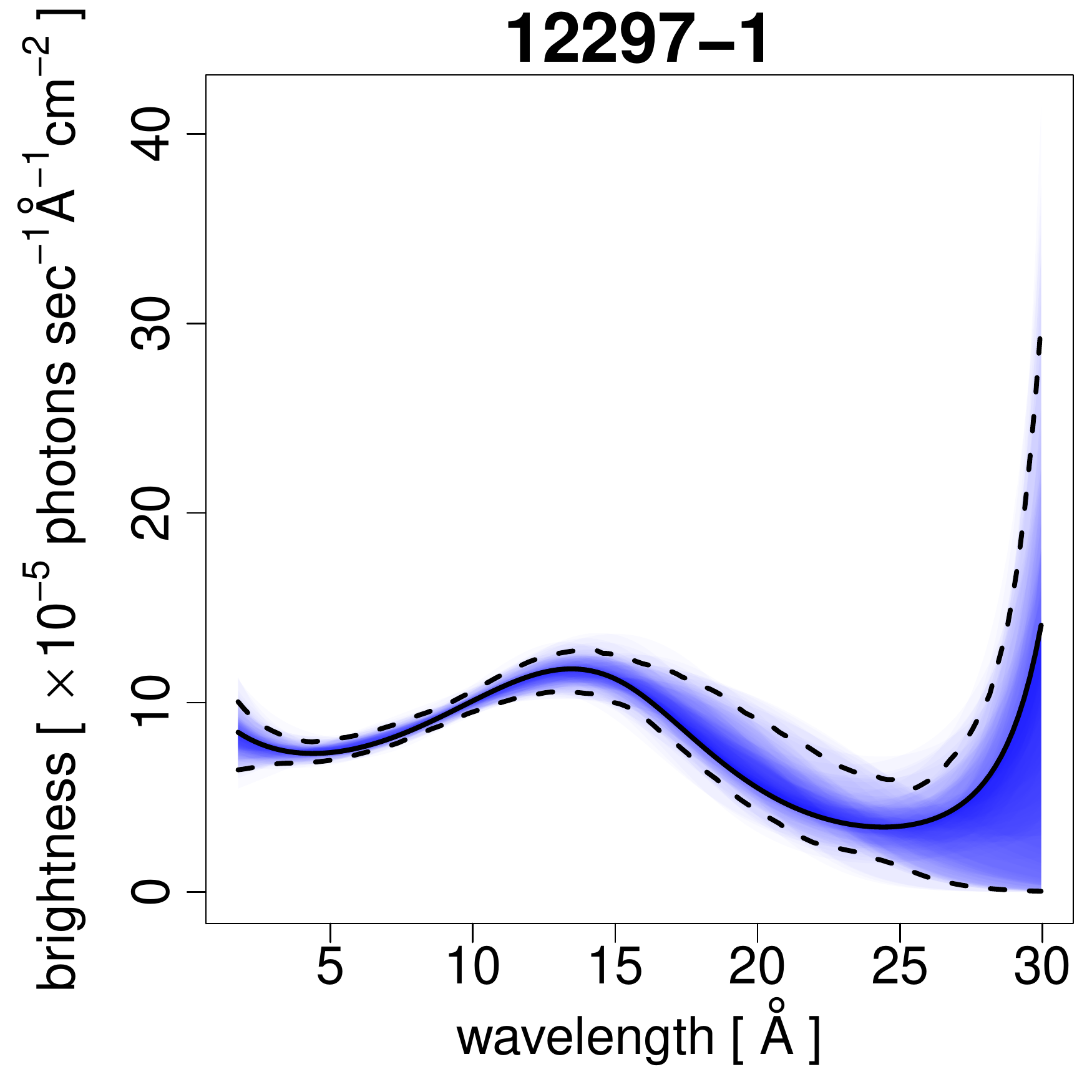}
\includegraphics[width=0.25\linewidth]{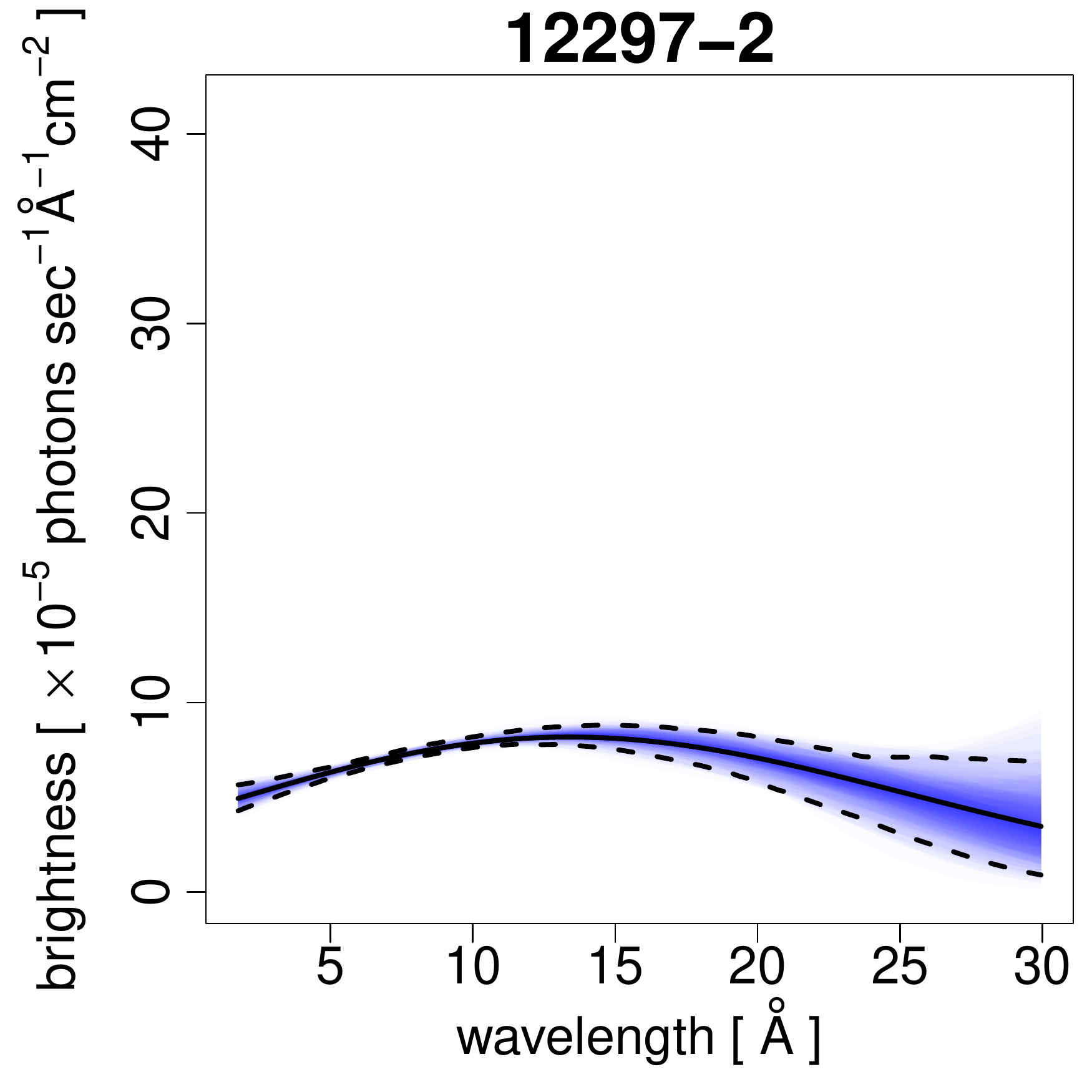}
\includegraphics[width=0.25\linewidth]{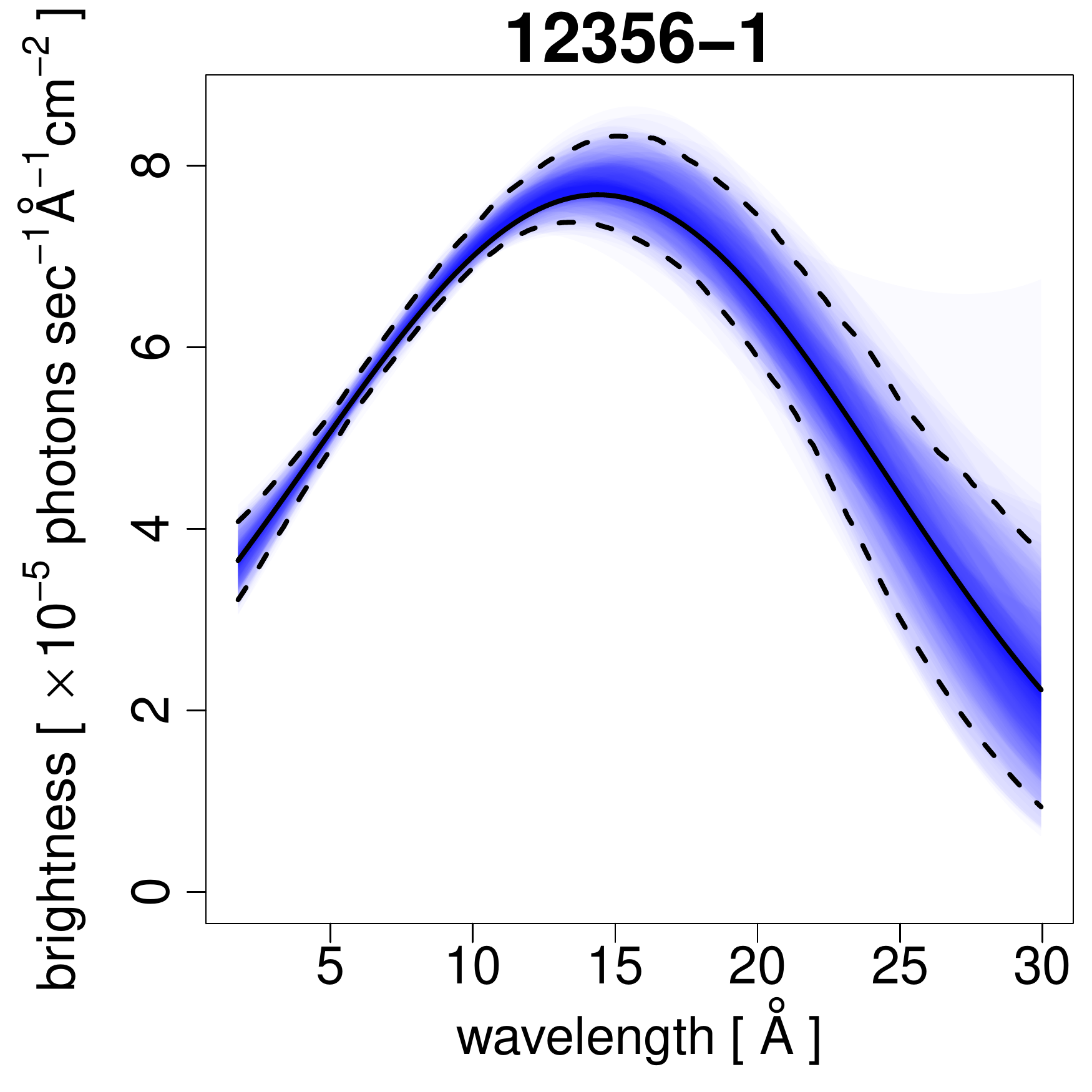}\\
\includegraphics[width=0.25\linewidth]{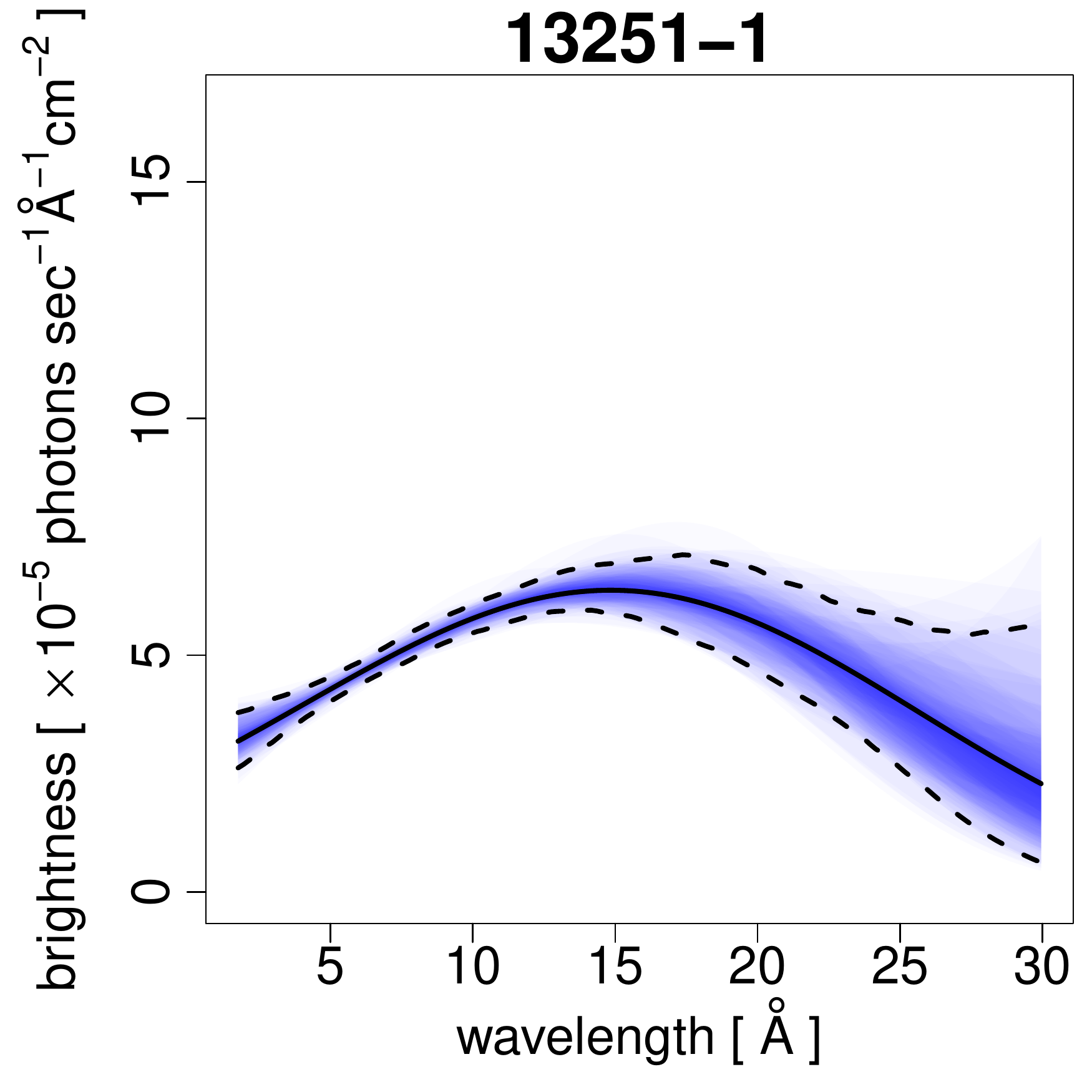}
\includegraphics[width=0.25\linewidth]{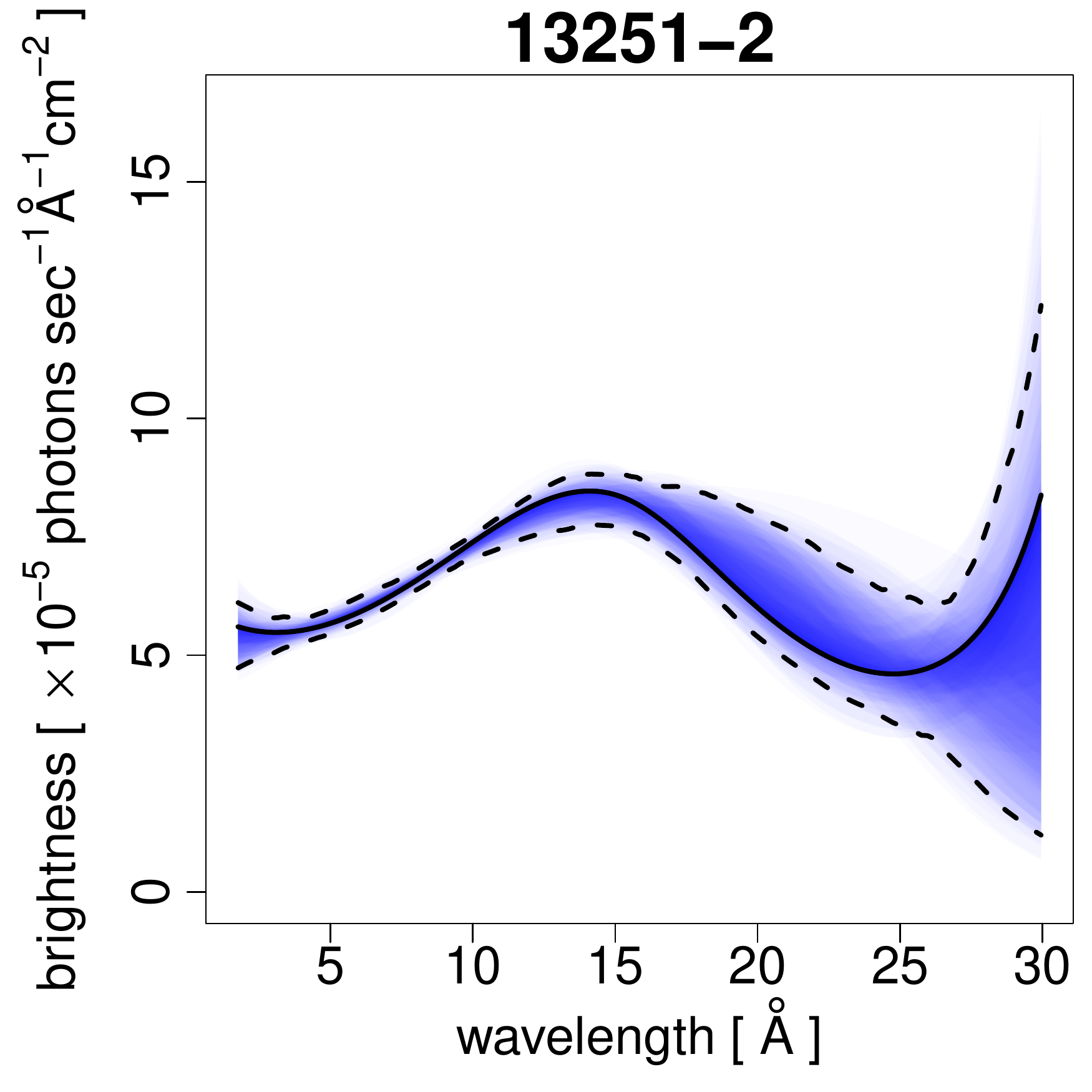}
\includegraphics[width=0.25\linewidth]{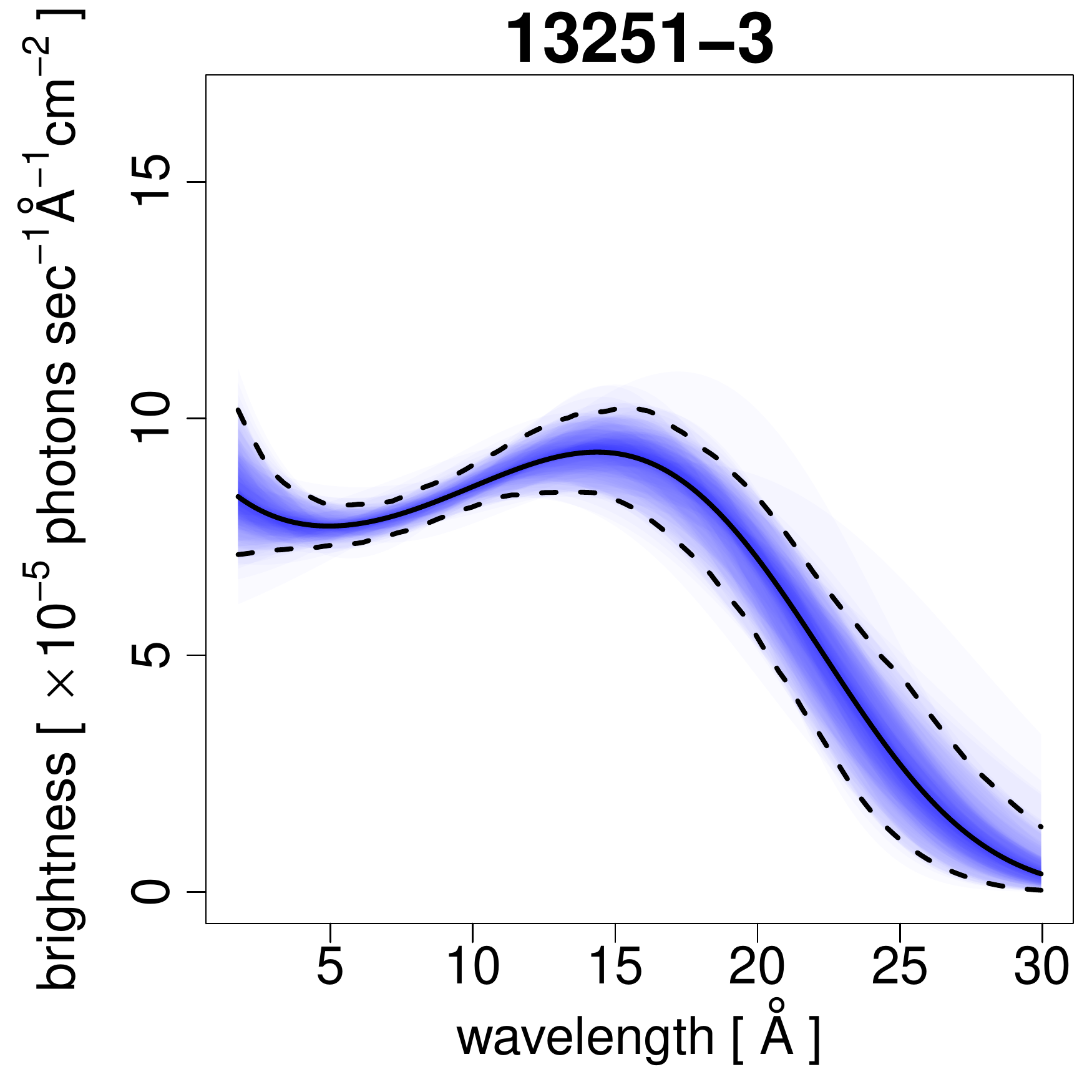}\\
\includegraphics[width=0.25\linewidth]{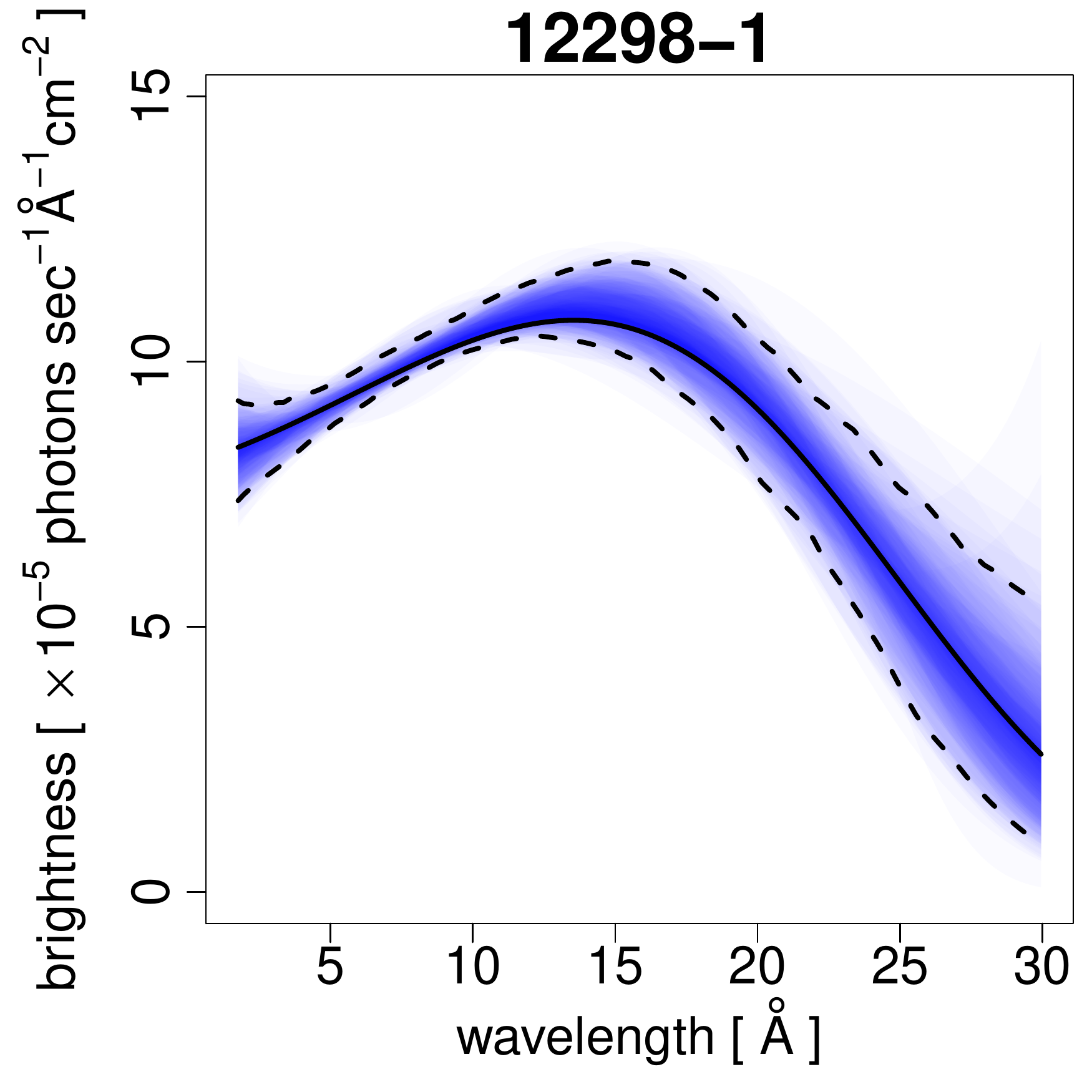}
\includegraphics[width=0.25\linewidth]{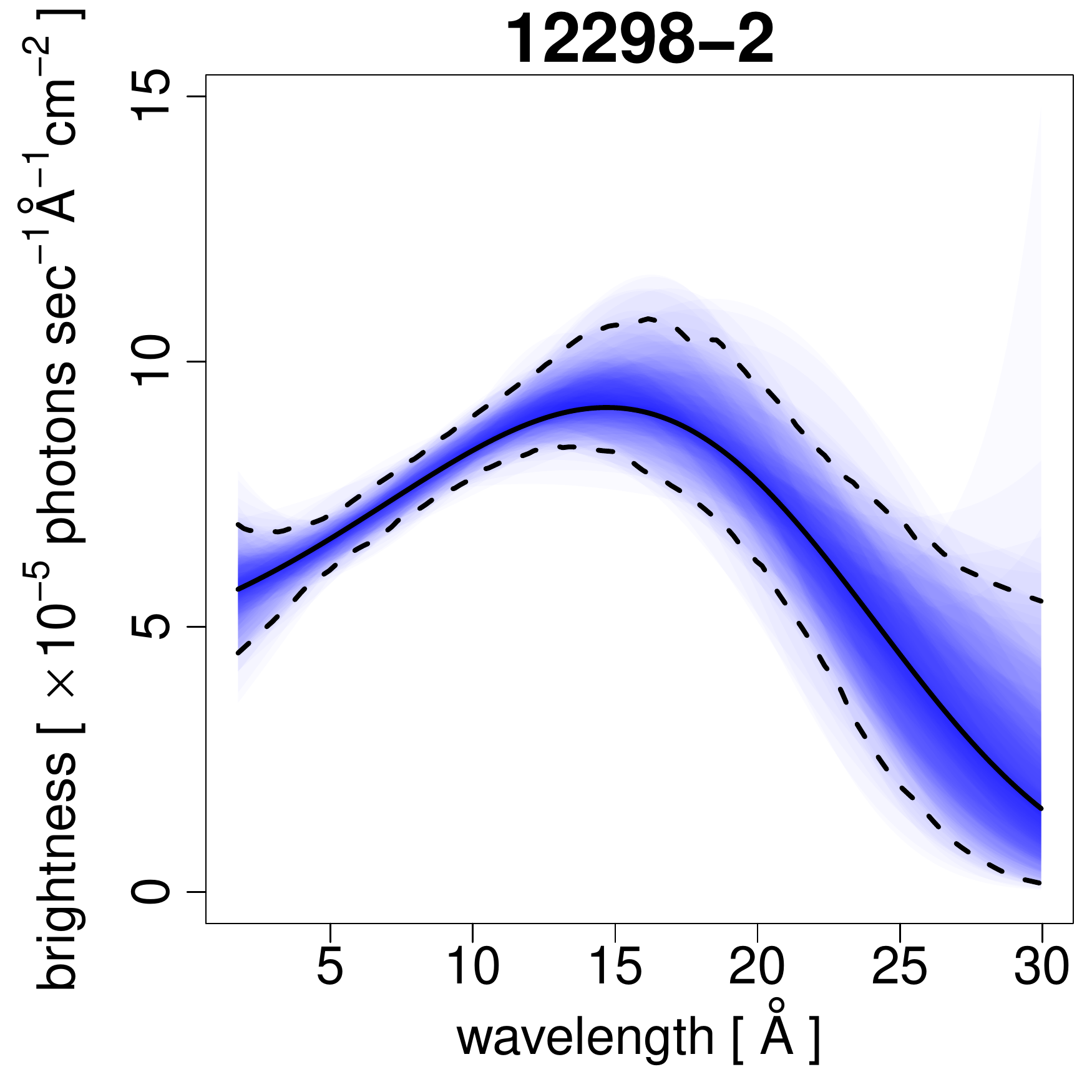} \hfill \\
\includegraphics[width=0.25\linewidth]{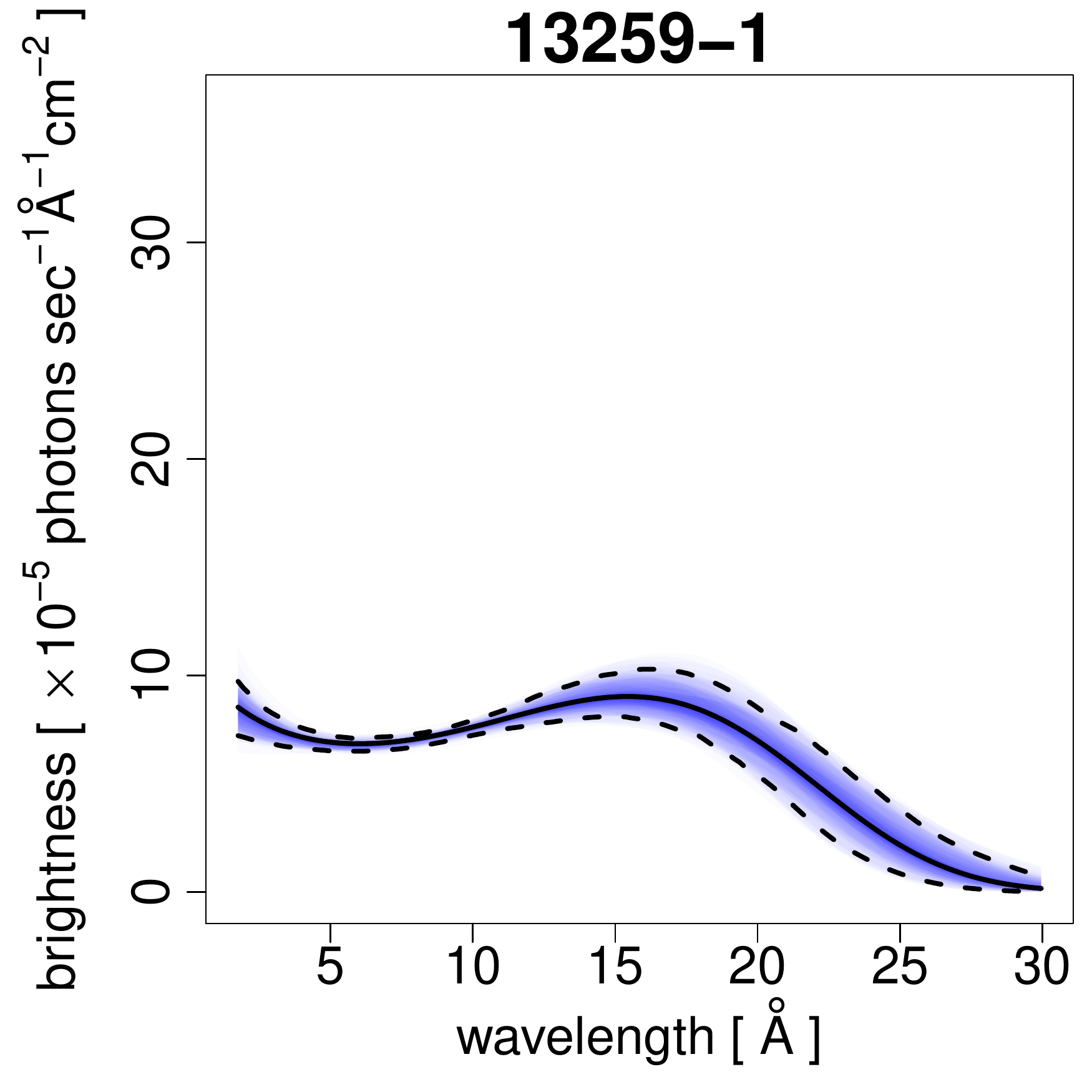}
\includegraphics[width=0.25\linewidth]{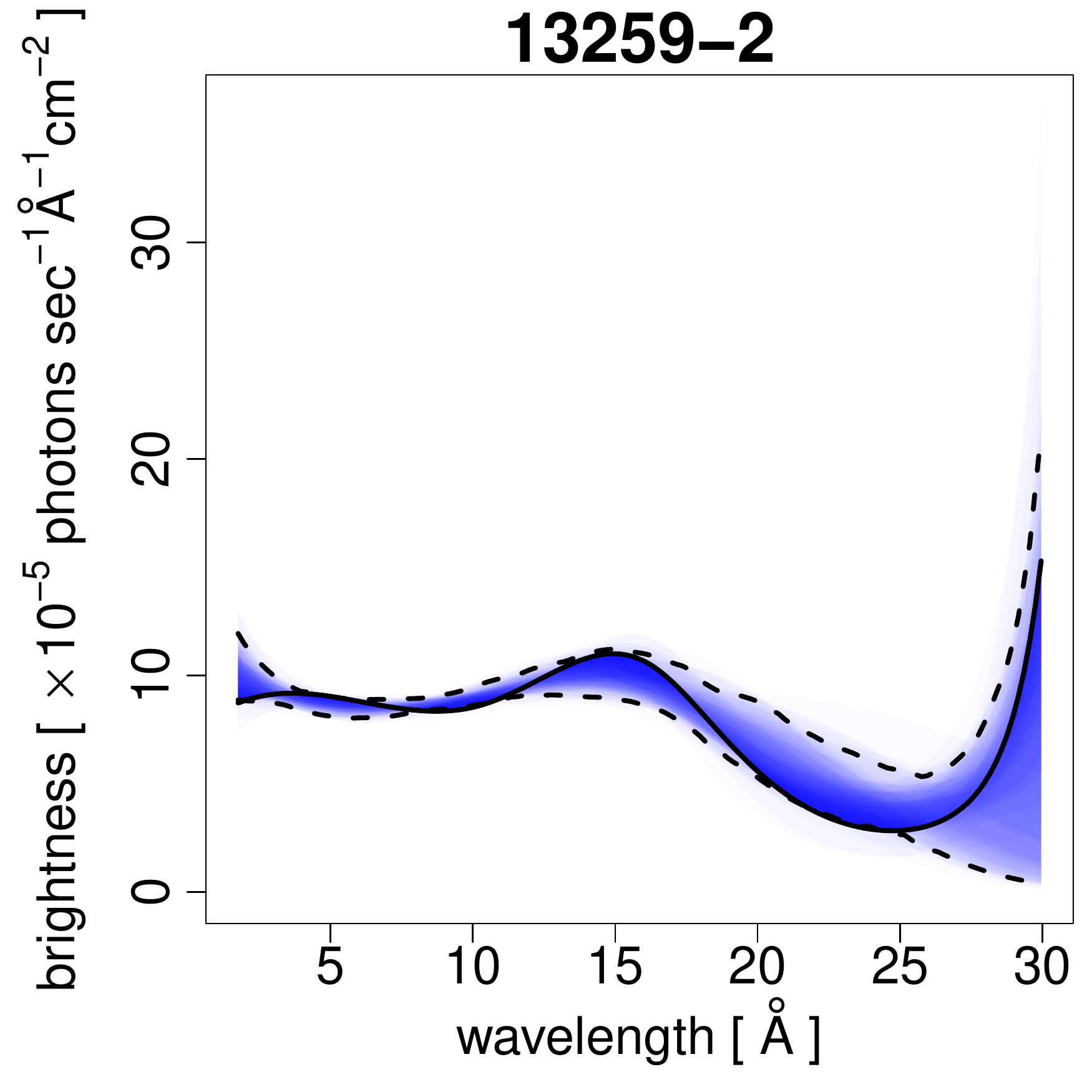}
\includegraphics[width=0.25\linewidth]{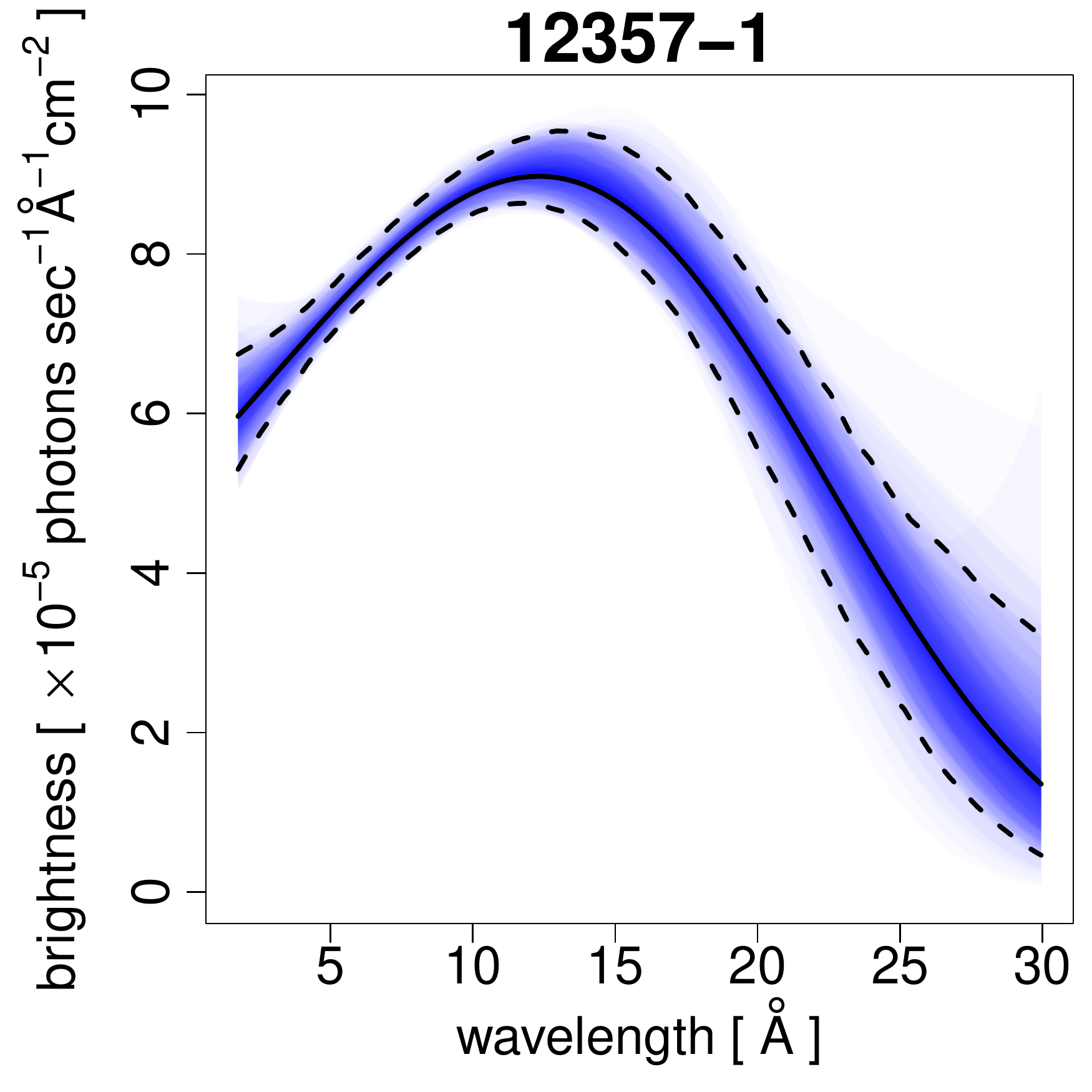}
\caption{\changes{Similar to Figure~\protect\ref{fig:cont} but for test functions 12297
  (77\%), 12356 (100\%), 13251 (69\%), 12298 (90\%), 13259 (62\%), and 12357 (100\%).
   Results are only plotted for simulations with $\hat B= B$ and
  $\hat{\bm{\pi}}=\bm{\pi}$; the proportions of simulations for which $\hat{B}=B$ and
$\hat{\bm{\pi}}=\bm{\pi}$ shown in the  parentheses.
Plots for the concatenated test function are not reported.
}
}
\label{fig:cont1}
\end{figure}

\section{Derivation of the MDL criterion~(\ref{eqn:mdl_break})}
\label{app:mdl2}
This appendix derives the MDL criterion~(\ref{eqn:mdl_break}).  As before, we need to calculate $\mathsf{CL}(\mathcal{D})=\mathsf{CL}(\hat{\mathcal{M}}) + \mathsf{CL}(\mathcal{D}|\hat{\mathcal{M}})$.  For model~(\ref{eqn:piece}), the codelength $\mathsf{CL}(\hat{\mathcal{M}})$ for any candidate model $\hat{\mathcal{M}}$ can be decomposed into
\[
  \mathsf{CL}(\hat{\mathcal{M}}) = 
\mathsf{CL}(B) + \mathsf{CL}(\bm{\pi}|B) + \mathsf{CL}(\bm{\hat{\Theta}}| B, \bm{\pi})
\]
Since $B$ is an integer, its codelength is
$\mathsf{CL}(B) = \log_2 B$.
Now for the codelength of $\bm{\pi}$.  Recall that $c_b$ is the length of the time segment $b$.  Therefore knowing the values of $(c_1, \ldots, c_B)$ is equivalent to knowing the values of all change points, $(\pi_1, \ldots, \pi_{B-1})$ .  Thus it suffices to encode $(c_1, \ldots, c_B)$, and because they are integers, we have
\[
  \mathsf{CL}(\bm{\pi}|B) = \sum^B_{b=1} \log_2 {p}_b.
\]
For the last term, we use the same arguments in Appendix~\ref{app:mdl1} and obtain
\[
  \mathsf{CL}(\bm{\hat{\Theta}}| B, \bm{\pi}) = 
\sum_{b=1}^{B}\mathsf{CL}(\bm{\hat{\theta}}_b| B, \bm{\pi}) = 
  \sum^{{B}}_{b=1}\left[\frac{1}{2}\|\hat{\bm{\theta}}_b\|_0\log_2 \{Nc(\bm\pi)\} + \log_2 {N \choose \|\hat{\bm{\eta}}_b\|_0}\right].
\]
Finally, $\mathsf{CL}(\mathcal{D}|\hat{\mathcal{M}})$ is given by the negative
of the log likelihood (base 2) of the candidate model being considered, which
gives
\[
  \mathsf{CL}(\mathcal{D}|\hat{\mathcal{M}})
=\sum^N_{i=1}\sum^J_{j=1} \sum^B_{b=1} I_b(t_j) l_{\mathrm{one}}\{ Y_{ij};\hat\boldtheta_b(B,\bm\pi)  \},
\]
with base 2 logarithm.  Changing the base of the logarithm terms, we
obtain~(\ref{eqn:mdl_break}).

\section{Supplement to Section \ref{sec:sim}}\label{app:sim}

Overlay plots of the fitted $\hat{f}$ to the 200 simulations generated with the test functions corresponding to ObsID~12297, 12298, 12356,12357, 13251, and 13259 are shown in Figure~\ref{fig:cont1}.
  These test functions share the same effective area curve depicted in Figure \ref{fig:cont}(d).

\bibliographystyle{rss}
\bibliography{raywongref}

\begin{thebibliography}{52}
\expandafter\ifx\csname natexlab\endcsname\relax\def\natexlab#1{#1}\fi
\expandafter\ifx\csname url\endcsname\relax
  \def\url#1{\texttt{#1}}\fi
\expandafter\ifx\csname urlprefix\endcsname\relax\def\urlprefix{URL }\fi

\bibitem[{Akman and Raftery(1986)}]{Akman-Raftery86}
Akman, V. and Raftery, A. (1986) Asymptotic inference for a change-point
  poisson process.
\newblock \emph{the Annals of Statistics},  1583--1590.

\bibitem[{Aue \emph{et~al.}(2014)Aue, Cheung, Lee and Zhong}]{Aue-et-al14b}
Aue, A., Cheung, R. C.~Y., Lee, T. C.~M. and Zhong, M. (2014) Segmented model
  selection in quantile regression using the minimum description length
  principle.
\newblock \emph{Journal of the American Statistical Association}.
\newblock To appear.

\bibitem[{Aue and Lee(2011)}]{Aue-Lee11}
Aue, A. and Lee, T. C.~M. (2011) {On image segmentation using information
  theoretic criteria}.
\newblock \emph{Annals of Statistics}, \textbf{39}, 2912--2935.

\bibitem[{Bopp and Stencel(1981)}]{bopp:sten:81}
Bopp, B. and Stencel, R. (1981) {The FK Comae stars}.
\newblock \emph{The Astrophysical Journal}, \textbf{247}, L131--L134.

\bibitem[{Buhmann(2003)}]{Buhmann03}
Buhmann, M.~D. (2003) \emph{Radial basis functions: theory and
  implementations}.
\newblock Cambridge: Cambridge university press.

\bibitem[{Carlin \emph{et~al.}(1992)Carlin, Gelfand and Smith}]{carl:etal:92}
Carlin, B.~P., Gelfand, A.~E. and Smith, A.~F. (1992) {Hierarchical Bayesian
  analysis of changepoint problems}.
\newblock \emph{Applied statistics},  389--405.

\bibitem[{Chan and Zhang(2007)}]{chan:zhan:07}
Chan, H.~P. and Zhang, N.~R. (2007) Scan statistics with weighted observations.
\newblock \emph{Journal of the American Statistical Association}, \textbf{102},
  595--602.

\bibitem[{Chen and Chen(2008)}]{Chen-Chen08}
Chen, J. and Chen, Z. (2008) {Extended Bayesian information criteria for model
  selection with large model spaces}.
\newblock \emph{Biometrika}, \textbf{95}, 759--771.

\bibitem[{Chib(1998)}]{chib:98}
Chib, S. (1998) Estimation and comparison of multiple change-point models.
\newblock \emph{Journal of econometrics}, \textbf{86}, 221--241.

\bibitem[{Cohen \emph{et~al.}(2010)Cohen, Drake, Kashyap, Korhonen, Elstner and
  Gombosi}]{Cohen-Drake-Kashyap10}
Cohen, O., Drake, J., Kashyap, V., Korhonen, H., Elstner, D. and Gombosi, T.
  (2010) Magnetic structure of rapidly rotating fk comae-type coronae.
\newblock \emph{The Astrophysical Journal}, \textbf{719}, 299--306.

\bibitem[{Davis and Yau(2013)}]{Davis-Yau13}
Davis, R.~A. and Yau, C.~Y. (2013) Consistency of minimum description length
  model selection for piecewise stationary time series models.
\newblock \emph{Electronic Journal of Statistics}, \textbf{7}, 381--411.

\bibitem[{Drake \emph{et~al.}(2008)Drake, Chung, Kashyap, Korhonen,
  Van~Ballegooijen and Elstner}]{Drake-Chung-Kashyap08}
Drake, J.~J., Chung, S.~M., Kashyap, V., Korhonen, H., Van~Ballegooijen, A. and
  Elstner, D. (2008) X-ray spectroscopic signatures of the extended corona of
  fk comae.
\newblock \emph{The Astrophysical Journal}, \textbf{679}, 1522--1530.

\bibitem[{Elstner and Korhonen(2005)}]{Elstner-Korhonen05}
Elstner, D. and Korhonen, H. (2005) Flip-flop phenomenon: observations and
  theory.
\newblock \emph{arXiv preprint astro-ph/0501343}, \textbf{326}, 278--282.

\bibitem[{Friedman \emph{et~al.}(2010)Friedman, Hastie and
  Tibshirani}]{Friedman-Hastie-Tibshirani10}
Friedman, J., Hastie, T. and Tibshirani, R. (2010) Regularization paths for
  generalized linear models via coordinate descent.
\newblock \emph{Journal of Statistical Software}, \textbf{33}, 1.

\bibitem[{Green(1995)}]{gree:95}
Green, P.~J. (1995) Reversible jump {M}arkov chain {M}onte {C}arlo computation
  and {B}ayesian model determination.
\newblock \emph{Biometrika}, \textbf{82}, 711--732.

\bibitem[{Gr{\"u}nwald \emph{et~al.}(2005)Gr{\"u}nwald, Myung and
  Pitt}]{grunwald2005advances}
Gr{\"u}nwald, P.~D., Myung, I.~J. and Pitt, M.~A. (2005) \emph{Advances in
  minimum description length: Theory and applications}.
\newblock MIT press.

\bibitem[{G{\"u}del(2004)}]{Gudel04}
G{\"u}del, M. (2004) X-ray astronomy of stellar coronae.
\newblock \emph{The Astronomy and Astrophysics Review}, \textbf{12}, 71--237.

\bibitem[{Kashyap \emph{et~al.}(2002)Kashyap, Drake, G{\"u}del and
  Audard}]{Kashyap-Drake-Gudel02}
Kashyap, V.~L., Drake, J.~J., G{\"u}del, M. and Audard, M. (2002) Flare heating
  in stellar coronae.
\newblock \emph{The Astrophysical Journal}, \textbf{580}, 1118--1132.

\bibitem[{Korhonen \emph{et~al.}(1999)Korhonen, Berdyugina, Hackman, Duemmler,
  Ilyin and Tuominen}]{korh:etal:99}
Korhonen, H., Berdyugina, S., Hackman, T., Duemmler, R., Ilyin, I. and
  Tuominen, I. (1999) Study of fk comae berenices. i. surface images for 1994
  and 1995.
\newblock \emph{Astronomy and Astrophysics}, \textbf{346}, 101--110.

\bibitem[{Lai and Xing(2011)}]{lai:xing:11}
Lai, T.~L. and Xing, H. (2011) {A simple Bayesian approach to multiple
  change-points}.
\newblock \emph{Statistica Sinica}, \textbf{21}, 539.

\bibitem[{Leclerc(1989)}]{Leclerc89}
Leclerc, Y.~G. (1989) Constructing simple stable descriptions for image
  partitioning.
\newblock \emph{International Journal of Computer Vision}, \textbf{3}, 73--102.

\bibitem[{Lee \emph{et~al.}(2011)Lee, Kashyap, van Dyk, Connors, Drake, Izem,
  Meng, Min, Park, Ratzlaff, Siemiginowska and Zezas}]{lee:etal:11}
Lee, H., Kashyap, V.~L., van Dyk, D.~A., Connors, A., Drake, J.~J., Izem, R.,
  Meng, X.~L., Min, S., Park, T., Ratzlaff, P., Siemiginowska, A. and Zezas, A.
  (2011) Accounting for calibration uncertainties in {X}-ray analysis:
  Effective areas in spectral fitting.
\newblock \emph{The Astrophysical Journal}, \textbf{731}, 126--144.

\bibitem[{Lee(1997)}]{Lee97c}
Lee, T. C.~M. (1997) \emph{Some Models and Methods in Image Segmentation}.
\newblock Ph.D. thesis, Macquarie University, Sydney, Australia.

\bibitem[{Lee(1998)}]{Lee98:segcor}
Lee, T. C.~M. (1998) Segmenting images corrupted by correlated noise.
\newblock \emph{IEEE Transactions on Pattern Analysis and Machine
  Intelligence}, \textbf{20}, 481--492.

\bibitem[{Lee(2002{\natexlab{a}})}]{Lee02}
Lee, T. C.~M. (2002{\natexlab{a}}) Automatic smoothing for discontinuous
  regression functions.
\newblock \emph{Statistica Sinica}, \textbf{12}, 823--842.

\bibitem[{Lee(2002{\natexlab{b}})}]{Lee02:genetic}
Lee, T. C.~M. (2002{\natexlab{b}}) On algorithms for ordinary least squares
  regression spline fitting: A comparative study.
\newblock \emph{Journal of Statistical Computation and Simulation},
  \textbf{72}, 647--663.

\bibitem[{Loader(1992)}]{Loader92}
Loader, C.~R. (1992) A log-linear model for a poisson process change point.
\newblock \emph{The Annals of Statistics},  1391--1411.

\bibitem[{MacKay(2003)}]{mack:03}
MacKay, D. J.~C. (2003) \emph{Information Theory, Inference, and Algorithms}.
\newblock Cambridge, UK: Cambridge University Press.

\bibitem[{McKee(2012)}]{mcke:12}
McKee, M. (2012) {Superflares' erupt on some Sun-like stars}.
\newblock \emph{Nature News,},  May 16, 2012.

\bibitem[{Mei \emph{et~al.}(2011)Mei, Han and Tsui}]{Mei-et-al11}
Mei, Y., Han, S.~W. and Tsui, K.-L. (2011) Early detection of a change in
  poisson rate after accounting for population size effects.
\newblock \emph{Statistica Sinica}, \textbf{21}, 597.

\bibitem[{Moreno \emph{et~al.}(2005)Moreno, Casella and
  Garcia-Ferrer}]{more:etal:05}
Moreno, E., Casella, G. and Garcia-Ferrer, A. (2005) {An objective Bayesian
  analysis of the change point problem}.
\newblock \emph{Stochastic Environmental Research and Risk Assessment},
  \textbf{19}, 191--204.

\bibitem[{Ninomiya(2014)}]{Ninomiya14}
Ninomiya, Y. (2014) Change-point model selection via {AIC}.
\newblock \emph{Annals of the Institute of Statistical Mathematics},  1--19.

\bibitem[{Park(2010)}]{park:10}
Park, J.~H. (2010) {Structural change in US presidents' use of force}.
\newblock \emph{American Journal of Political Science}, \textbf{54}, 766--782.

\bibitem[{Park \emph{et~al.}(2006)Park, Kashyap, Siemiginowska, van Dyk, Zezas,
  Heinke and Wargelin}]{park:etal:06}
Park, T., Kashyap, V., Siemiginowska, A., van Dyk, D.~A., Zezas, A., Heinke, C.
  and Wargelin, B.~J. (2006) {B}ayesian estimation of hardness ratios: Modeling
  and computations.
\newblock \emph{The Astrophysical Journal}, \textbf{652}, 610--628.

\bibitem[{Park \emph{et~al.}(2012)Park, Krafty and S{\'a}nchez}]{park:etal:12}
Park, T., Krafty, R.~T. and S{\'a}nchez, A.~I. (2012) {Bayesian semi-parametric
  analysis of Poisson change-point regression models: application to
  policy-making in Cali, Colombia}.
\newblock \emph{Journal of applied statistics}, \textbf{39}, 2285--2298.

\bibitem[{Raftery and Akman(1986)}]{raft:akma:86}
Raftery, A. and Akman, V. (1986) {Bayesian analysis of a Poisson process with a
  change-point}.
\newblock \emph{Biometrika},  85--89.

\bibitem[{Rissanen(1989)}]{Rissanen89}
Rissanen, J. (1989) \emph{Stochastic Complexity in Statistical Inquiry}.
\newblock World Scientific, Singapore.

\bibitem[{Rissanen(2007)}]{Rissanen07}
Rissanen, J. (2007) \emph{Information and Complexity in Statistical Modeling}.
\newblock Springer.

\bibitem[{Ruppert \emph{et~al.}(2003)Ruppert, Wand and
  Carroll}]{Ruppert-Wand-Carroll03}
Ruppert, D., Wand, M.~P. and Carroll, R.~J. (2003) \emph{Semiparametric
  regression}.
\newblock Cambridge: Cambridge university press.

\bibitem[{Scargle(1998)}]{scar:98}
Scargle, J.~D. (1998) Studies in astronomical time series analysis. {V.
  Bayesian} blocks, a new method to analyze structure in photon counting data.
\newblock \emph{The Astrophysical Journal}, \textbf{504}, 405--418.

\bibitem[{Scargle \emph{et~al.}(2013)Scargle, Norris, Jackson and
  Chiang}]{scar:etal:13}
Scargle, J.~D., Norris, J.~P., Jackson, B. and Chiang, J. (2013) Studies in
  astronomical time series analysis. {VI. Bayesian} block representations.
\newblock \emph{The Astrophysical Journal}, \textbf{764}, 167--192.

\bibitem[{Shen and Zhang(2012)}]{shen:zhan:12}
Shen, J.~J. and Zhang, N.~R. (2012) {Change-point model on nonhomogeneous
  Poisson processes with application in copy number profiling by
  next-generation DNA sequencing}.
\newblock \emph{The Annals of Applied Statistics}, \textbf{6}, 476--496.

\bibitem[{Strassmeier(2009)}]{stras:09}
Strassmeier, K.~G. (2009) Starspots.
\newblock \emph{The Astronomy and Astrophysics Review}, \textbf{17}, 251--308.

\bibitem[{Tibshirani(1996)}]{Tibshirani96}
Tibshirani, R. (1996) Regression shrinkage and selection via the lasso.
\newblock \emph{Journal of the Royal Statistical Society: Series B},
  \textbf{58}, 267--288.

\bibitem[{van Dyk \emph{et~al.}(2006)van Dyk, Connors, Esch, Freeman, Kang,
  Karovska, Kashyap, Siemiginowska and Zezas}]{Dyk-Connors-Esch06}
van Dyk, D.~A., Connors, A., Esch, D.~N., Freeman, P., Kang, H., Karovska, M.,
  Kashyap, V., Siemiginowska, A. and Zezas, A. (2006) Deconvolution in
  high-energy astrophysics: Science, instrumentation, and methods.
\newblock \emph{Bayesian Analysis}, \textbf{1}, 189--236.

\bibitem[{van Dyk \emph{et~al.}(2001)van Dyk, Connors, Kashyap and
  Siemiginowska}]{vand:etal:01}
van Dyk, D.~A., Connors, A., Kashyap, V. and Siemiginowska, A. (2001) Analysis
  of energy spectra with low photon counts via {B}ayesian posterior simulation.
\newblock \emph{The Astrophysical Journal}, \textbf{548}, 224--243.

\bibitem[{van Dyk and Kang(2004)}]{vand:kang:04}
van Dyk, D.~A. and Kang, H. (2004) Highly structured models for spectral
  analysis in high-energy astrophysics.
\newblock \emph{Statistical Science}, \textbf{19}, 275--293.

\bibitem[{Worsley(1986)}]{Worsley86}
Worsley, K.~J. (1986) Confidence regions and tests for a change-point in a
  sequence of exponential family random variables.
\newblock \emph{Biometrika}, \textbf{73}, 91--104.

\bibitem[{Xu \emph{et~al.}(2014)Xu, van Dyk, Kashyap, Siemiginowska, Connors,
  Drake, Meng, Ratzlaff and Yu}]{xu:etal:14}
Xu, J., van Dyk, D.~A., Kashyap, V.~L., Siemiginowska, A., Connors, A., Drake,
  J.~J., Meng, X.~L., Ratzlaff, P. and Yu, Y. (2014) A fully {B}ayesian method
  for jointly fitting instrumental calibration and x-ray spectral models.
\newblock \emph{The Astrophysical Journal}, \textbf{794}, 97 (21pages).

\bibitem[{Yao(1988)}]{Yao88}
Yao, Y.-C. (1988) Estimating the number of change-points via {S}chwarz'
  criterion.
\newblock \emph{Statistics and Probability Letters}, \textbf{6}, 181--189.

\bibitem[{Yau \emph{et~al.}(2015)Yau, Tang and Lee}]{Yau-et-al15}
Yau, C.-Y., Tang, C.-M. and Lee, T. C.~M. (2015) Estimation of multiple-regime
  threshold autoregressive models with structural breaks.
\newblock \emph{Journal of the American Statistical Association}, \textbf{110},
  1175--1186.

\bibitem[{Zhang and Siegmund(2007)}]{Zhang-Siegmund07}
Zhang, N.~R. and Siegmund, D.~O. (2007) {A modified Bayes information criterion
  with applications to the analysis of comparative genomic hybridization data}.
\newblock \emph{Biometrics}, \textbf{63}, 22--32.

\end{thebibliography}

\end{document}